\pdfoutput=1
\documentclass[12pt]{article}
\usepackage[parfill]{parskip}
\usepackage{graphicx}
\usepackage{numprint}
\npthousandsep{,}
\usepackage[nomessages]{fp}
\usepackage{MnSymbol}
\usepackage{tabulary}
\usepackage{multirow}
\usepackage{url}
\usepackage{times}
\usepackage{xspace}
\usepackage{lscape}
\usepackage{enumitem}
\usepackage[top=2.45cm,bottom=2.45cm,left=3.17cm,right=3.17cm,twoside]{geometry}
\usepackage{chngcntr}
\usepackage{natbib}
\bibpunct{(}{)}{;}{a}{,}{,}
\usepackage[nottoc]{tocbibind}
\settocbibname{References}
\usepackage[table]{xcolor}
\usepackage[nodayofweek]{datetime}
\usepackage{setspace}
\usepackage{authblk}
\usepackage{lineno}
\usepackage[group-separator={,}]{siunitx}
\usepackage{hyphenat}
\usepackage{tikz}
\usepackage{fmtcount}
\DeclareSIUnit{\molar}{M}
\usepackage{subfig}
\usepackage{hyphenat}
\usepackage{seqsplit}
\usepackage[font={footnotesize }]{caption}

\renewcommand{\seqinsert}{\ifmmode\allowbreak\else\-\fi}
  
\newcommand{\isobamwolmil}{4.5}

\newcommand{\isobamwolpercent}{2.5}

\newcommand{\corrmin}{0.61}

\newcommand{\corrpvalmax}{< 2.2e-16}

\newcommand{\corrminrep}{0.94}

\newcommand{\nbCone}{1033}

\newcommand{\percentCone}{86.4}

\newcommand{\nbCtwo}{162}

\newcommand{\percentCtwo}{13.6}

\newcommand{\nbCtwoa}{103}

\newcommand{\nbCtwob}{59}

\newcommand{\DEisoglm}{80}

\newcommand{\percentDEisoglm}{6.7}

\newcommand{\DEisoglmclassone}{5}

\newcommand{\DEisoglmup}{75}

\newcommand{\percentDEisoglmup}{93.8}

\newcommand{\DEisoglmclasstwo}{57}

\newcommand{\DEisoglmclassthree}{22}

\newcommand{\DEisoglmCone}{6}

\newcommand{\DEisoglmCtwo}{74}

\newcommand{\percentDEisoglmCtwo}{92.5}

\newcommand{\DEisoglmCtwob}{52}

\newcommand{\DEisoglmCtwobpercent}{70.3}

\newcommand{\DEisoglmCtwoa}{22}

\newcommand{\DEisoglmCtwoapercent}{29.7}

\newcommand{\DEImfall}{41}

\newcommand{\DEImffive}{35}

\newcommand{\DEImfupFem}{7}

\newcommand{\DEImfallGLM}{28}

\newcommand{\percentDEImfall}{68.3}

\newcommand{\percentDEImfallGLM}{35}

\newcommand{\DEall}{93}

\newcommand{\DEallpercent}{7.8}

\newcommand{\DEallnowMel}{7}

\newcommand{\DEallwMel}{86}

\newcommand{\DEallwMelMelanogaster}{3}

\newcommand{\DEallwMelMelanogasterpercent}{3.5}

\newcommand{\DEallwMelArthropod}{30}

\newcommand{\DEallwMelArthropodpercent}{34.9}

\newcommand{\DEallwMelEcdysozoan}{53}

\newcommand{\DEallwMelEcdysozoanpercent}{61.6}

\newcommand{\chisq}{4.82}

\newcommand{\chisqpVal}{0.09}

\newcommand{\minGeneStage}{948}

\newcommand{\minGeneStagepercent}{79.3}

\newcommand{\medianGene}{1110}

\newcommand{\minGeneStagetwoCount}{856}

\newcommand{\minGeneStagetwoCountpercent}{71.6}

\newcommand{\minreadsIsomil}{0.1}

\newcommand{\minreadsIsopercent}{0.3}

\newcommand{\maxreadsIsomil}{7}

\newcommand{\maxreadsIsopercent}{8.5}

\newcommand{\medianreadsIsomil}{1.7}

\newcommand{\medianreadsIsopercent}{1.6}

\title{Dynamics of \textit{Wolbachia pipientis} gene expression across the \textit{Drosophila melanogaster} life cycle.\vspace{4ex}}
\author[1]{\normalsize  Florence~Gutzwiller}
\author[2]{\normalsize  Catarina~R.~Carmo}
\author[3,4]{\normalsize  Danny~E.~Miller}
\author[5]{\normalsize  Danny~W.~Rice}
\author[5]{\normalsize  Irene~L.G.~Newton}
\author[3,4]{\normalsize  R.~Scott~Hawley}
\author[2]{\normalsize  Luis~Teixeira}
\author[1]{\normalsize  Casey~M.~Bergman\vspace{4ex}}
\affil[1]{\footnotesize Faculty of Life Sciences, University of Manchester, Manchester, United Kingdom, M13 9PT.}
\affil[2]{\footnotesize Instituto Gulbenkian de Ci\^{e}ncia, Oeiras, Portugal}
\affil[3]{\footnotesize Stowers Institute for Medical Research, Kansas City, MO, USA, 64110}
\affil[4]{\footnotesize Department of Molecular and Integrative Physiology, University of Kansas Medical Center, Kansas City, KS, USA, 66160}
\affil[5]{\footnotesize Department of Biology, Indiana University, Bloomington, Indiana, USA\vspace{2ex}}

\begin{document}

\begin{titlepage}
\maketitle
Author for correspondence: \\
Casey M. Bergman \\
Faculty of Life Sciences \\
University of Manchester \\
Michael Smith Building \\
Oxford Road \\
Manchester, M13 9PT \\ 
United Kingdom \\
casey.bergman@manchester.ac.uk \\
\end{titlepage}
\clearpage

\section*{Abstract}
\label{sec:Abstract}
\doublespacing


Symbiotic interactions between microbes and their multicellular hosts have manifold impacts on molecular, cellular and organismal biology. To identify candidate bacterial genes involved in maintaining endosymbiotic associations with insect hosts, we analyzed genome-wide patterns of gene expression in the $\alpha$-proteobacteria \textit{Wolbachia pipientis} across the life cycle of \textit{Drosophila melanogaster} using public data from the modENCODE project that was generated in a \textit{Wolbachia}-infected version of the ISO1 reference strain. We find that the majority of \textit{Wolbachia} genes are expressed at detectable levels in \textit{D. melanogaster} across the entire life cycle, but that only \DEallpercent\% of 1195 \textit{Wolbachia} genes exhibit robust stage- or sex-specific expression differences when studied in the ``holo-organism" context. \textit{Wolbachia} genes that are differentially expressed during development are typically up-regulated after \textit{D. melanogaster} embryogenesis, and include many bacterial membrane, secretion system and ankyrin-repeat containing proteins. Sex-biased genes are often organized as small operons of uncharacterized genes and are mainly up-regulated in adult \textit{Drosophila} males in an age-dependent manner suggesting a potential role in cytoplasmic incompatibility. Our results indicate that large changes in \textit{Wolbachia} gene expression across the \textit{Drosophila} life-cycle are relatively rare when assayed across all host tissues, but that candidate genes to understand host-microbe interaction in facultative endosymbionts can be successfully identified using holo-organism expression profiling. Our work also shows that mining public gene expression data in \textit{D. melanogaster} provides a rich set of resources to probe the functional basis of the \textit{Wolbachia}-\textit{Drosophila} symbiosis and annotate the transcriptional outputs of the \textit{Wolbachia} genome.

\textbf{Keywords:} \textit{Wolbachia}, \textit{Drosophila}, gene expression, symbiosis, development

\clearpage
\section*{Background}
\label{sec:Background}
\doublespacing

Intracellular bacterial symbioses provide powerful systems to investigate the diverse consequences of coevolution between microbes and their hosts. Some bacterial endosymbiotic interactions are beneficial to both organisms, resulting in coadapations that generate mutual dependency. These obligate symbioses are often characterized by ancient phylogenetic associations, restriction of microbes to specialized host cells, provision of essential nutrients from microbe to host, and microbial genome reduction (reviewed in \citet{dale_molecular_2006,moran_genomics_2008}). Other microbial endosymbiotic interactions are obligate for the microbe, but are non-essential (``facultative") from the standpoint of the host. Facultative bacterial symbioses are typically more short-lived evolutionarily and microbial effects on hosts are not always beneficial \citep{dale_molecular_2006,moran_genomics_2008}. For example, the $\alpha$-proteobacteria \textit{Wolbachia pipientis} is often found in facultative associations with many arthropod species, inducing a range of host effects from reproductive parasitism to viral protection (reviewed in \citet{zug_bad_2015}). In fact, even within a single host species such as \textit{Drosophila melanogaster}, closely-related variants of the same \textit{Wolbachia} lineage can lead to either mutualistic or pathogenic interactions \citep{min_wolbachia_1997,chrostek_wolbachia_2013}. Facultative endosymbionts are of particular interest since some may represent a transitional state between free-living bacteria and obligate mutualists, thus offering insights into both the early evolutionary stages of mutualism and the propagation of invasive pathogens \citep{dale_molecular_2006, moran_functional_2006}.

Efforts to identify microbial genes that maintain infections of facultative endosymbionts are hampered by the inability to culture and manipulate these species in a free-living state. Likewise, the lack of extreme genome reduction in facultative endosymbionts does not allow the mere existence of a gene to provide \textit{prima facie} evidence of its importance in a particular host context, as it does in mutualist species with highly reduced genomes \citep{moran_functional_2006}. Therefore, candidate genes in facultative endosymbionts that might mediate interaction with their hosts have been primarily identified using comparative genomic approaches. For example, sequencing of the first \textit{Wolbachia} genome from \textit{D. melanogaster} (\textit{w}Mel) revealed an unusually large number ankyrin repeat domain (ANK) encoding genes relative to other bacteria \citep{wu_phylogenomics_2004}. Large numbers of ANK-containing genes are also observed in the genomes of other \textit{Wolbachia}  strains that form facultative associations with arthropod hosts \citep{iturbe-ormaetxe_distribution_2005,duron_variability_2007,siozios_diversity_2013}, while few ANK-containing genes are found in the obligate \textit{Wolbachia} endosymbionts of nematodes \citep{foster_wolbachia_2005,darby_analysis_2012}. Comparative genomic analysis of more closely-related strains of \textit{Wolbachia} has also been used to identify candidate genes involved in host-symbiont interaction \citep{iturbe-ormaetxe_distribution_2005,sinkins_wolbachia_2005,duron_variability_2007,chrostek_wolbachia_2013,woolfit_genomic_2013}. For example, a cluster of eight genes (called the Octomom region), identified as being specifically amplified in the pathogenic ``Popcorn" (\textit{w}MelPop) strain of \textit{Wolbachia} from \textit{D. melanogaster} \citep{chrostek_wolbachia_2013,woolfit_genomic_2013}, was recently shown to cause the high bacterial titres and virulence associated with this strain \citep{chrostek_mutualism_2015}.

Genome-wide gene expression profiling offers an alternative approach to identify candidate genes involved in host-symbiont interactions. Both transcriptomic and proteomic approaches have been used to study how bacterial gene regulation changes in native host tissues for obligate endosymbionts \citep{wilcox_consequences_2003,moran_regulation_2005,reymond_different_2006,bennuru_stage-specific_2011,darby_analysis_2012,rao_effects_2012,luck_concurrent_2014}. However, genome-wide expression profiling has not yet been used extensively to study gene expression dynamics for facultative endosymbionts in their native host context \citep{slatko_wolbachia_2014}. Recently, \citet{darby_integrated_2014} conducted transcriptomic and proteomic analysis of a \textit{Wolbachia} strain from \textit{D. melanogaster} (\textit{w}MelPop-CLA) and \citet{baldridge_proteomic_2014} profiled the proteome of \textit{Wolbachia} wStr from the planthopper \textit{Laodelphax striatellus}, both stably transinfected into non-native host cell lines from the mosquito \textit{Aedes albopictus}. Likewise, two recent studies have used \textit{w}MelPop-CLA transinfected in non-native \textit{Aedes} cell lines to identify small non-coding RNAs (ncRNAs) by high-throughput sequencing \citep{mayoral_wolbachia_2014,woolfit_discovery_2015}. \citet{woolfit_discovery_2015} also generated transcriptomic data from two \textit{Wolbachia} strains (\textit{w}MelPop and \textit{w}MelCS) in native host tissues (heads of \textit{D. melanogaster}), but did not attempt to identify differentially-expressed genes that may be involved in host-microbe interactions.

Here we report global gene expression dynamics for a facultative endosymbiont across the life cycle of a native arthropod host, taking advantage of a previously uncharacterized \textit{Wolbachia} infection in the reference strain that was used for the \textit{D. melanogaster} genome project. We show that the \textit{D. melanogaster} ISO1 reference strain was originally infected with \textit{Wolbachia} prior to being donated to the \textit{Drosophila} stock center, where after it was used by the modENCODE project to generate deep total RNA-seq data from 30 time points across the \textit{D. melanogaster} life cycle including embryos, larvae, pupae, adult males, and adult females \citep{graveley_developmental_2011,brown_diversity_2014,duff_genome-wide_2015}. Using this rich transcriptomic resource, we show that the majority of \textit{Wolbachia} genes are expressed across the life cycle, but that most \textit{Wolbachia} genes show stable expression across different host stages and sexes when studied at the whole-fly level. We identify a set of 80 genes that show reproducible changes in expression levels in at least one life-cycle stage, the majority of which are up-regulated after embryonic development with peaks of expression in early larval, late pupal or adult stages. We also identify 41 genes that show expression differences between males and females, with the majority of these sex-biased genes being up-regulated in males and showing age-dependent effects. Genes with stage- or sex-specific expression differences include chaperones, ANK-containing genes, and genes with predicted membrane or secretion system function, but most have no known function. Our results provide general insight into the dynamics of gene expression in a facultative endosymbiont across life-cycle stages and sexes of an arthropod host, and provide a set of resources to further explore the functional basis of the model \textit{Wolbachia}-\textit{Drosophila} symbiosis.

\clearpage

\section*{Results and Discussion}
\label{Results and Discussion}

\subsection*{The \textit{D. melanogaster} ISO1 reference strain is infected with \textit{Wolbachia}}

As a control for another project, we obtained the ISO1 reference strain \citep{brizuela_genetic_1994} used for the \textit{D. melanogaster} genome project from the Bloomington \textit{Drosophila} Stock Center (BDSC) and sequenced its genome. We discovered that the BDSC ISO1 sample contained a large number of \textit{Wolbachia} sequences (\isobamwolmil\ million reads, \isobamwolpercent\% of total) when mapped against a ``holo-genome" comprised of the \textit{D. melanogaster} plus \textit{W. pipientis} \textit{w}Mel reference genomes \citep{adams_genome_2000,wu_phylogenomics_2004}. The observation of \textit{Wolbachia} sequences in the ISO1 stock was unexpected, since at no point since its original sequencing by the Berkeley \textit{Drosophila} Genome Project (BDGP) and Celera Genomics in 2000 had \textit{Wolbachia} sequences been reported in this strain \citep{adams_genome_2000,celniker_finishing_2002,hoskins_release_2015}. In fact, direct searches of assembled or unassembled ISO1 sequences from the BDGP failed to detect any evidence of \textit{Wolbachia} \citep{wu_phylogenomics_2004,salzberg_serendipitous_2005}. 

We investigated the cause of the discrepancy of \textit{Wolbachia} sequences in the BDGP and BDSC ISO1 genomic data by first establishing the provenance of these lineages. As documented in their database, the BDSC ISO1 sub-strain was donated directly to the stock center by Jim Kennison in 1994. The ISO1 sub-strain used by the BDGP was obtained from Gerry Rubin's lab in the late 1990's (Roger Hoskins, personal communication). The Rubin lab ISO1 sub-strain was obtained in the mid-1990s from Jim Kennison \textit{via} at least one intermediate lab (Todd Laverty, personal communication). Thus, in contrast to naive assumptions, the BDSC ISO1 sub-strain is neither a direct descendant nor progenitor of the ISO1 sub-strain used in the \textit{D. melanogaster} genome project, and these two sub-strains have had independent trajectories since at least 1994.

The presence of \textit{Wolbachia} sequences in our genomic data, but not in BDGP genomic data, could be explained in a number of ways. For example, the BDGP ISO1 sub-strain could have lost its infection, or \textit{Wolbachia} sequences could have been eliminated during nuclear enrichment steps that were used to reduce mitochondrial DNA in the original Sanger sequencing libraries (Roger Hoskins, personal communication) \citep{adams_genome_2000}. Though less likely because horizontal transfer of \textit{Wolbachia} has not been observed in \textit{D. melanogaster} but spontaneous loss is relatively common \citep{hoffmann_population_1998,richardson_population_2012}, the ISO1 sub-strain could alternatively have acquired a \textit{Wolbachia} infection in the BDSC, where many stocks are known to be infected \citep{clark_widespread_2005}. To resolve when and where the BDSC ISO1 sub-strains lost (or acquired) its \textit{Wolbachia} infection, we obtained ISO1 sub-strains from three additional sources: (i) the original sub-strain, maintained continuously by Jim Kennison since its synthesis in 1986; (ii) the sub-strain used for the \textit{Drosophila} genome project, maintained by the BDGP since the late 1990's; and (iii) a sub-strain maintained by Gerry Rubin's lab since the mid-1990's, which was the progenitor of the BDGP sub-strain. The Rubin lab sub-strain was never treated for a \textit{Wolbachia} infection using antibiotics (Gerry Rubin, personal communication). We performed diagnostic PCR for \textit{Wolbachia} on all four ISO1 sub-strains before and after tetracycline treatment. We found that the ISO1 sub-strains from Jim Kennison's lab and the BDSC were infected with \textit{Wolbachia}, but those from the Rubin lab and the BDGP were not (data not shown). Amplification of \textit{Wolbachia} sequences was only observed in flies before antibiotic treatment, indicating that the source of \textit{Wolbachia} DNA sequences is not due to nuclear integration, which occurs rarely in \textit{D. melanogaster} \citep{huang_natural_2014}. We conclude that the original ISO1 strain synthesized by Kennison was infected with \textit{Wolbachia}, that an infected ISO1 sub-strain was donated by Kennison to the BDSC in 1994, and that an ISO1 sub-strain that was cured of (or spontaneously lost) its infection was obtained by the Rubin lab and donated to the BDGP. These results also show that, in addition to a high proportion of lab stocks carrying \textit{Wolbachia} \citep{clark_widespread_2005}, sub-lines of the same \textit{D. melanogaster} stock may vary in their \textit{Wolbachia} infection status.

\begin{figure}[h!] \centering 
\includegraphics[width=0.7\textwidth]{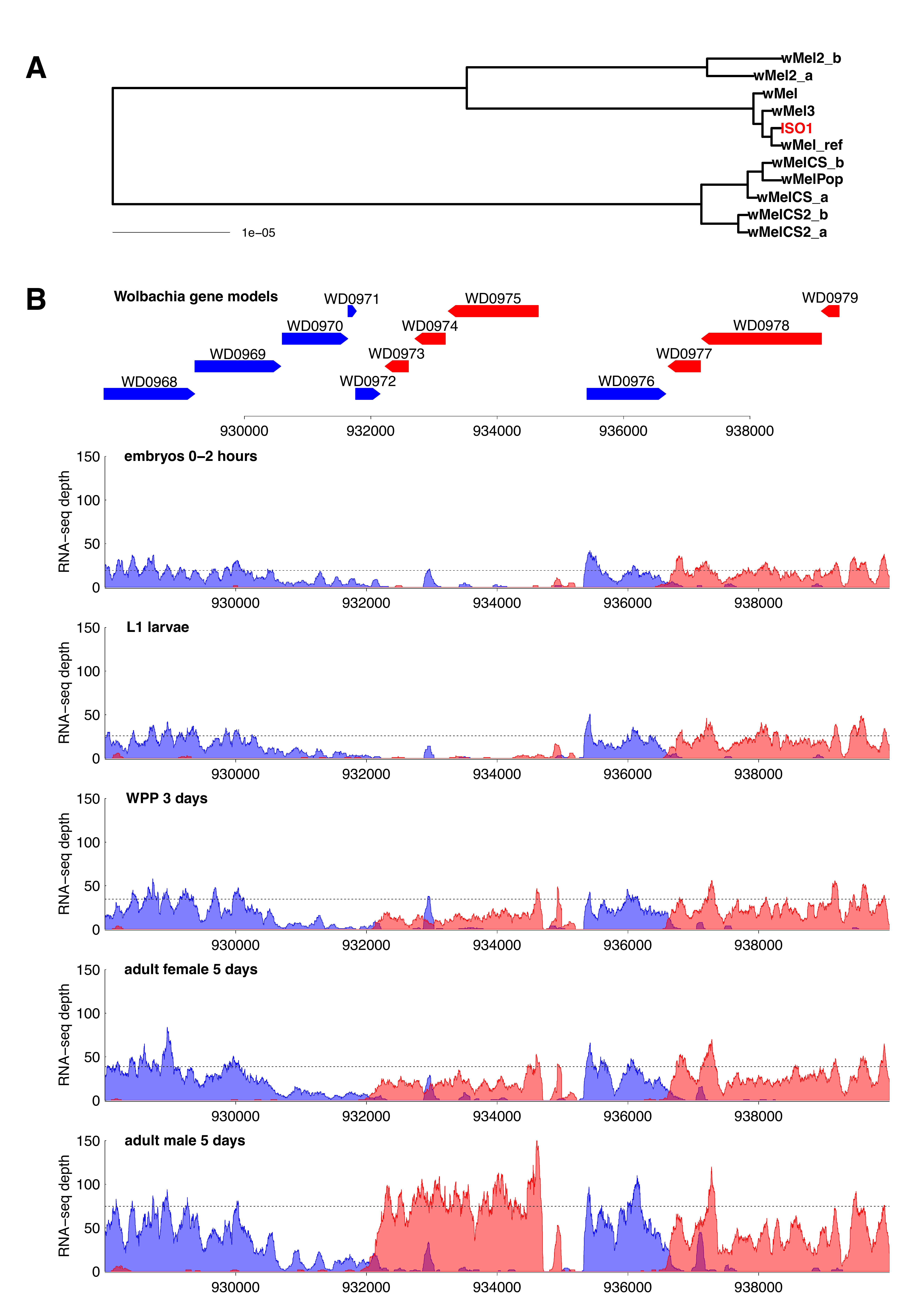} 
\caption{{\bf Phylogeny and expression of \textit{Wolbachia} in \textit{D. melanogaster}.} A. Phylogenetic tree of \textit{Wolbachia} strains based on whole genome sequences from this study (ISO1, red), \citet{wu_phylogenomics_2004} (\textit{w}Mel\_ref), \citet{chrostek_wolbachia_2013} (all others). The scale bar for branch lengths is units of substitutions per site. The \textit{Wolbachia} variant in ISO1 is very closely related to the \textit{w}Mel reference genome and to the \textit{w}Mel variant in \citet{chrostek_wolbachia_2013}. B. Gene models and RNA-seq coverage plots for a 12-gene window of the \textit{Wolbachia} genome showing gene expression levels in representative stages of the \textit{D. melanogaster} modENCODE RNA-seq life-cycle time course. Gene models (pointed rectangles) and RNA-seq coverage (strand-specific wiggle plots of number of reads mapped to each base pair) are shown on the forward and reverse strands in blue and red, respectively. RNA-seq plots are shown on the same absolute y-axis scale. To provide an internal normalization factor for comparison across samples, mean coverage of the stably-expressed Wsp/WD1063 gene (not shown in this interval) divided by twenty is depicted by the dashed line in each panel. A unannotated non-coding RNA transcript is express anti-sense to the 3' end of WD0974. This example shows a set of three consecutive genes (WD0973, WD0974 and WD0975) that are specifically up-regulated in males in comparison to neighboring genes and co-transcribed as a single operon.} 
\label{fig:combined_figure} 
\end{figure}

After confirming that the BDSC ISO1 sub-strain is indeed infected with \textit{Wolbachia}, we next addressed which of the major lineages of \textit{Wolbachia} segregating in \textit{D. melanogaster} infects this strain \citep{richardson_population_2012,chrostek_wolbachia_2013,woolfit_genomic_2013}. To do this, we assembled a consensus sequence from ISO1 reads that mapped to the \textit{w}Mel reference and generated a whole-genome phylogeny jointly with the \textit{w}Mel reference genome \citep{wu_phylogenomics_2004} and genomes from known \textit{Wolbachia} genotypes \citep{chrostek_wolbachia_2013}. This analysis showed that the \textit{Wolbachia} infection in the BDSC ISO1 sub-strain is from a \textit{w}Mel-like genotype that is very closely related to both the \textit{w}Mel reference genome sequence \citep{wu_phylogenomics_2004} and the \textit{w}Mel type strain recently reported by \citet{chrostek_wolbachia_2013} (Figure \ref{fig:combined_figure}A). The very close relationship between the \textit{w}Mel genotype in ISO1 and the \textit{w}Mel reference genome allows functional genomic data collected in ISO1 to be easily and accurately mapped to the reference genome sequence, and for reference genome annotations to closely reflect the content of the ISO1 \textit{Wolbachia} genome.

\subsection*{The modENCODE developmental time-course reveals global stability in \textit{Wolbachia} gene expression across the \textit{D. melanogaster} life cycle}

Establishing that the BDSC ISO1 sub-strain is infected with \textit{Wolbachia} is important since this strain is widely used in \textit{Drosophila} genomics, including being one of the strains used by modENCODE to profile the transcriptome of \textit{D. melanogaster} \citep{graveley_developmental_2011,brown_diversity_2014,duff_genome-wide_2015}. In particular, modENCODE generated total RNA-seq libraries from BDSC ISO1 that span 30 time points across the \textit{D. melanogaster} life cycle including multiple stages from embryos, larvae, pupae, and both adult sexes, with two biological replicates being available for 24 of the 30 time points \citep{graveley_developmental_2011,brown_diversity_2014,duff_genome-wide_2015}. We emphasize that the modENCODE  total RNA-seq libraries were made from whole organisms, and thus any tissue-specific differences in \textit{Wolbachia} gene expression will be averaged, as will sex-specific differences in non-adult stages. We tested whether the \textit{Wolbachia} infection in ISO1 could be detected in modENCODE total RNA-seq libraries by mapping reads  to the combined \textit{D. melanogaster} plus \textit{W. pipientis} holo-genome reference. We found that the modENCODE total RNA-seq libraries do contain large numbers of \textit{Wolbachia} sequences, with a median of \medianreadsIsomil\ million reads per sample (range: \minreadsIsomil-\maxreadsIsomil\ million) mapping to the \textit{Wolbachia} genome, corresponding to a median of \medianreadsIsopercent\% (range: \minreadsIsopercent\%-\maxreadsIsopercent\%) of the total number of RNA-seq reads mapped in each sample (Supplemental File \ref{itm:supplemental_file_1.tsv}). As shown in Figure \ref{fig:combined_figure}B, coverage and strand-specificity of the modENCODE RNA-seq dataset is high enough to show clear correspondence with the boundaries of the majority of annotated \textit{Wolbachia} gene models, given their presumed operonic structure and lack of annotated untranslated regions.  These results further confirm that the BDSC ISO1 sub-strain is indeed infected with \textit{Wolbachia}, allowing (and also requiring) the modENCODE RNA-seq developmental time course to be analyzed in the context of the \textit{Wolbachia}-\textit{Drosophila} symbiosis.

We next counted reads and estimated expression levels in transcripts per million (TPM) for each of the 1195 characterized protein-coding genes in the \textit{Wolbachia} genome in each of the modENCODE RNA-seq samples (Supplemental File \ref{itm:supplemental_file_2.tsv}). At least \minGeneStage\ (\minGeneStagepercent\%) \textit{Wolbachia} genes were expressed (defined as having non-zero TPM estimates) across all stages and replicates (Supplemental File \ref{itm:supplemental_file_1.tsv}). Using the more stringent definition of $\geq2$ mapped reads per gene used by \citet{darby_integrated_2014}, more \textit{Wolbachia} genes can be detected as being expressed in each sample from the modENCODE whole-organism RNA-seq data set ($>$\minGeneStagetwoCount\ genes, $>$\minGeneStagetwoCountpercent\% of total) than could be detected by these authors using combined transcriptomic and proteomic profiling of \textit{Wolbachia} \textit{w}MelPop-CLA expression in cell culture (66.8\%). The two samples with the lowest percentage of expressed genes were also among those with the fewest total reads, and both had many more expressed genes in the biological replicate from the same stage (Embryos 22-24 hr: 948 vs 1111, Embryos 12-14 hr: 994 vs 1112, Supplemental File \ref{itm:supplemental_file_1.tsv}), indicating they may be outliers with respect to the actual number of genes expressed because of low sequencing coverage. The median number of genes expressed (non-zero TPM) across all samples is \medianGene, indicating that the majority of the \textit{Wolbachia} genome is transcriptionally active throughout the \textit{D. melanogaster} life cycle. 

\begin{figure}[ht] \centering
\includegraphics[width=1.0\textwidth]{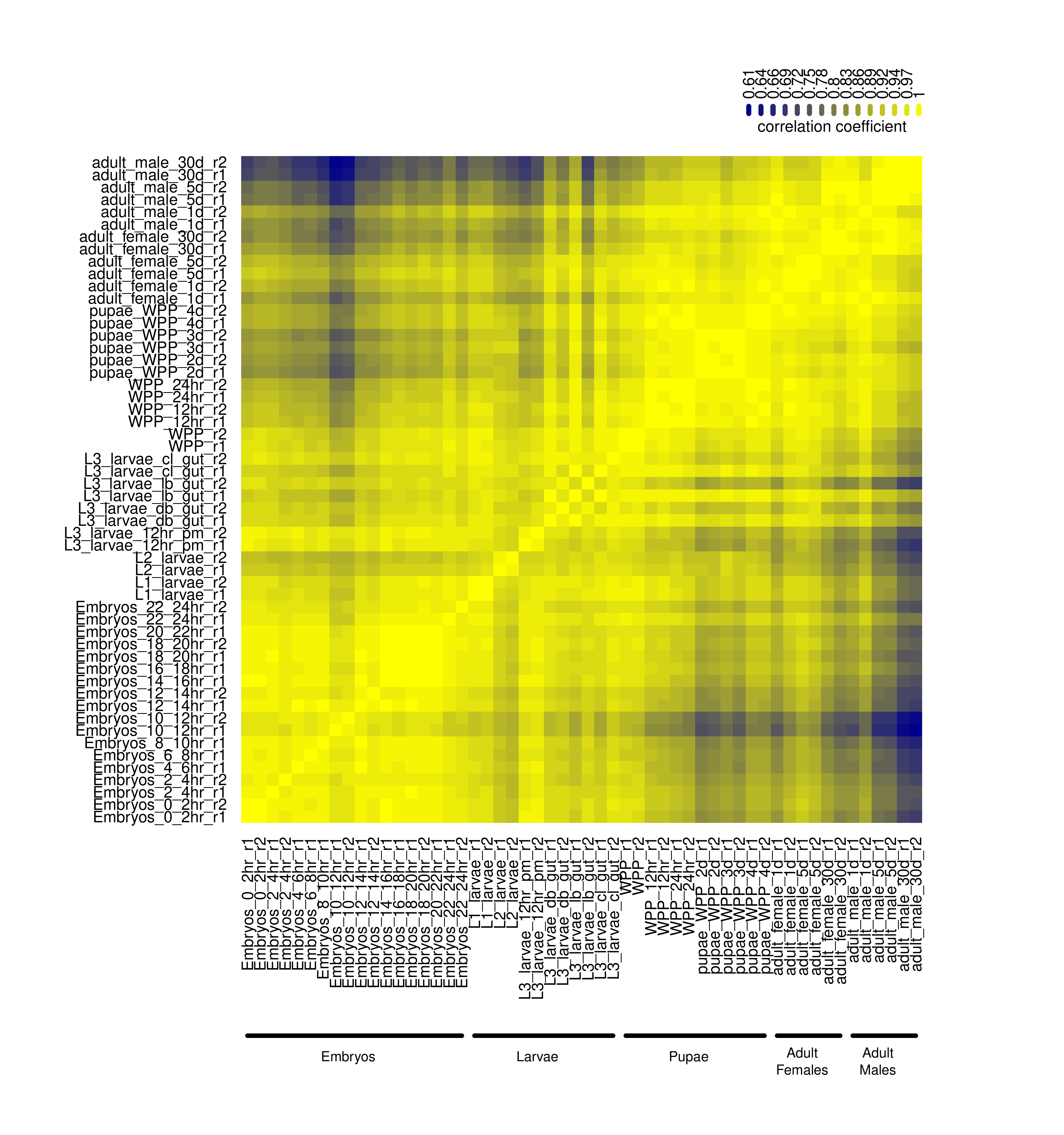} 
\caption{{\bf \textit{Wolbachia} gene expression levels are highly-correlated across the \textit{D. melanogaster} life cycle.} Each cell in the heatmap represents a Pearson correlation coefficient for a pair of samples in the ISO1 total RNA-seq dataset computed using expression levels in units of TPM for all genes. Higher similarity among pairs of samples is represented by bright yellow and lower similarity by dark blue, with all samples showing correlation coefficients of $>$\corrminrep. All but six stages in the modENCODE total RNA-seq time course have biological replicates (Embryos 4-6 hrs, Embryos 6-8 hrs, Embryos 8-10 hrs, Embryos 14-16 hrs, Embryos 16-18 hrs, and Embryos 20-22 hrs). Replicate samples from the same stage were collected in two independent series, denoted by \_r1 and \_r2 suffixes. Correlation among biological replicates of the same stage is very high, with the exception of late larval L3 stages (dark blue gut, light blue gut and clear gut) where samples from different stages of the same replicate series had higher correlation than replicate samples from the same stage, leading to the observed checkerboard pattern.} 
\label{fig:correlation_heatmap} 
\end{figure}

To provide an initial view of how \textit{Wolbachia} gene expression changes across the \textit{D. melanogaster} life-cycle, we measured the correlation of expression levels (in TPMs) across all \textit{Wolbachia} genes between all pairs of samples in the modENCODE total RNA-seq time course (Figure \ref{fig:correlation_heatmap}). Correlation among biological replicates of the same stage is very high ($r>\corrminrep$) and highly significant ($p\corrpvalmax$). The lowest correlation among biological replicates is observed for three stages from the third larval instar (L3 dark blue gut, L3 light blue gut and L3 clear gut), which show higher similarity between stages from the same replicate series than they do between biological replicates from the same stage. Correlation of expression levels is also reasonably high and highly significant among all life-cycle stages ($r>\corrmin$, $p\corrpvalmax$). However, two weakly-differentiated, partially-overlapping clusters can be observed that span embryonic to white prepupal (WPP) stages, and late larval to adult stages, respectively (square blocks of yellow spanning multiple stages in Figure \ref{fig:correlation_heatmap}). Stage-specific clusters for embryonic 10-12 hours, larval L1, larval L2, and larval L3 samples, respectively, can also be observed in addition to the larger embryonic/pupal and pupal/adult clusters. Overall, these results suggest that differences in \textit{Wolbachia} gene expression do exist among \textit{D. melanogaster} life-cycle stages, but that the global pattern of \textit{Wolbachia} gene expression does not change dramatically across the \textit{D. melanogaster} life cycle when assayed at the level of the whole organism. 

Given the availability of an essentially-complete \textit{Wolbachia} developmental transcriptome, we can now ask if gene expression levels can be predicted based on the codon usage bias of protein-coding genes in a facultative endosymbiont like \textit{Wolbachia}, as has been assumed in the past \citep{wu_phylogenomics_2004}. We find that \textit{Wolbachia} shows no significant correlation between levels of transcription measured in transcripts per million (TPM) and effective number of codons ($N_{c}$) at any \textit{D. melanogaster} life-cycle stage (Supplemental Figure \ref{fig:codon_bias}). Thus, despite global stability in expression across the host life-cycle, \textit{Wolbachia} genes have not optimized their codon usage bias to reflect levels of transcription, nor does it appear that \textit{Wolbachia} gene expression levels are globally adapted to any particular host life-cycle stage. These results are consistent with the lack of association between codon bias and expression levels in other pathogenic and obligate endosymbionts \citep{andersson_codon_1996,palacios_strong_2002,herbeck_gene_2003,rispe_mutational_2004}, and may reflect the inefficacy of selection on codon usage in the low effective population sizes expected in endosymbionts, or the reduced number of tRNA genes in the \textit{Wolbachia} genome \citep{chan_gtrnadb:_2009}. Alternatively, this result could imply that \textit{Wolbachia} uses extensive post-transcriptional regulation to modulate protein expression levels, as has recently been reported in \textit{Buchnera} \citep{hansen_widespread_2014}.

\subsection*{A small subset of \textit{Wolbachia} genes show dynamic expression across the \textit{D. melanogaster} life cycle}

\begin{figure}[ht] \centering
\includegraphics[width=0.85\textwidth]{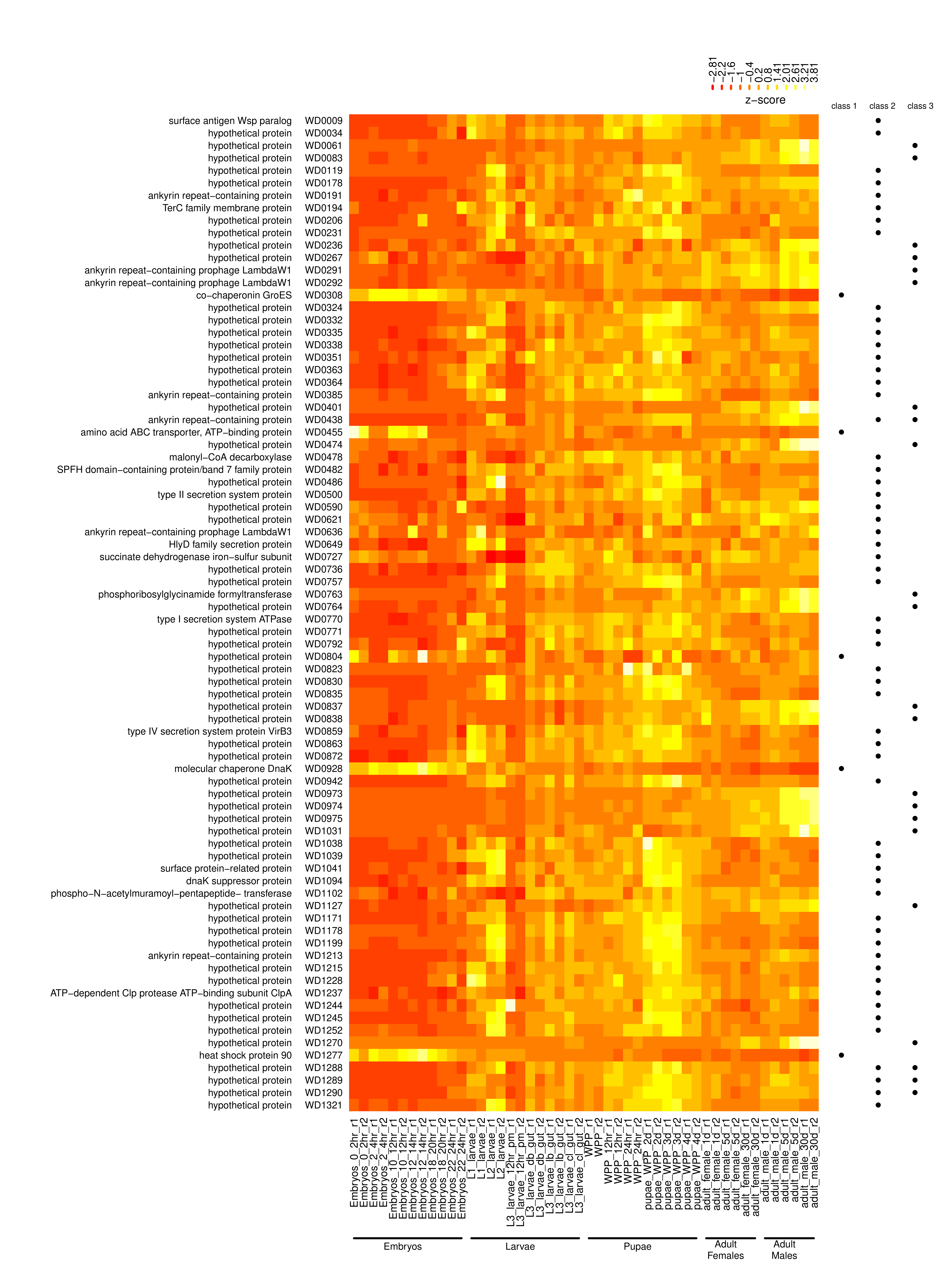}
\caption{{\bf \textit{Wolbachia} genes show differential expression across the \textit{D. melanogaster} life-cycle.} Row-normalized expression levels are visualized as a heatmap where each row represents a gene (ordered top-to-bottom by its position in the genome) and each cell represents the relative expression level for a particular sample in terms of Z-scores (observed TPM minus row mean TPM, divided by the standard deviation of TPMs for that row). Values higher than row means are represented by yellow, and values lower than row means are represented by red.  Gene names and identifiers are shown on the left. Membership in dynamically-expressed gene classes is shown by dots on the right. Class 1 includes genes that show down-regulation after embryogenesis. Class 2 includes genes that show up-regulation after embryogenesis, with peaks of expression in larval and pupal stages. Class 3 includes genes that show up-regulation after embryogenesis, with peaks of expression in adult. Classification of gene sets is not mutually exclusive. Stages that lack biological replicates in the modENCODE total RNA-seq time course were not used in this analysis and are not shown here. A version of this figure with non-row-normalized expression levels can be found in Supplemental Figure \ref{fig:heatmap_iso1DE_nonorm}.}
\label{fig:heatmap_iso1DE} 
\end{figure}

We next used the 24 stages in the modENCODE time-course that have biological replicates to identify \textit{Wolbachia} genes whose expression varies in a reproducible manner across the \textit{D. melanogaster} life cycle. Because of the large number of life-cycle stages and their complex developmental dependencies, an analysis involving all-by-all pairwise comparisons was deemed unfeasible. Instead, we used an omnibus test of changes in expression across all stages simultaneously using an ANOVA-like GLM approach \citep{mccarthy_differential_2012}. This analysis revealed a small subset of \textit{Wolbachia} genes (\DEisoglm/1195, \percentDEisoglm\%) that are differentially expressed in one or more life-cycle stage at an adjusted p-value of less than 0.05 (Figure \ref{fig:heatmap_iso1DE}, Supplemental File \ref{itm:supplemental_file_2.tsv}). All \DEisoglm\ genes have a greater than two-fold change between at least one pair of stages. The vast majority of \textit{Wolbachia} genes identified as differentially expressed across the \textit{D. melanogaster} life cycle in the modENCODE time-course show a common pattern of being expressed at lower relative levels in embryos (\DEisoglmup/\DEisoglm, \percentDEisoglmup\%), with higher expression in either larval, pupal and/or adult stages, and a transient decrease in expression at larval L3 (12 hrs). However, \numberstringnum{\DEisoglmclassone}\ genes show the opposite pattern of having higher relative expression in embryos with down-regulation later in the life cycle. The dynamics of up-regulated and down-regulated genes show nearly complementary transitions at the end of embryogenesis, suggesting response to common signals or possible cross-talk between these gene sets. We note that both up- and down-regulated genes exhibit a wide range of absolute expression levels, and many show quantitative shifts rather than dramatic qualitative changes in expression level (Supplemental Figure \ref{fig:heatmap_iso1DE_nonorm}).

To support conclusions about the global pattern of \textit{Wolbachia} gene expression dynamics across the \textit{D. melanogaster} life cycle based on differential expression analysis, we also performed probabilistic clustering on the entire modENCODE RNA-seq time course (including stages without replicates) \citep{si_model-based_2014} .  This analysis identified two main clusters that could be matched across independent clustering runs (Supplemental Figure \ref{fig:clustering}A). The first cluster contains the majority of \textit{Wolbachia} genes (\nbCone, \percentCone\%) and shows a pattern of relatively stable expression levels across the life cycle (Supplemental Figure \ref{fig:clustering}B). The second cluster contains the remaining \nbCtwo\ genes (\percentCtwo\%) and includes the vast majority of genes identified as differentially expressed in the life-cycle GLM (\DEisoglmCtwo/\DEisoglm, \percentDEisoglmCtwo\%). Only \numberstringnum{\DEisoglmCone}\ genes (GroES/WD0308, ABC transporter/WD0455, succinate dehydrogenase subunit/WD0727, WD0804, DnaK/WD0928, Hsp90/WD1277) identified as differentially expressed in the life-cycle GLM were not found in cluster 2 and are instead merged together with the stably-expressed genes in cluster 1, most of which (with the exception of succinate dehydrogenase subunit/WD0727) are all in the small, down-regulated class of differentially-expressed genes. 

Closer inspection of the number of runs in which a gene was assigned to cluster 2 revealed two distinct sub-clusters (Supplemental Figure \ref{fig:clustering}A), both of which show up-regulation after embryogenesis. The first, which we call cluster 2a, shows modest up-regulation after embryogenesis with the characteristic dip in expression at larval L3 (12 hrs) for a set of \nbCtwoa\ genes, and includes a substantial proportion (\DEisoglmCtwoa/\DEisoglmup, \DEisoglmCtwoapercent\%) of the up-regulated genes identified in the life-cycle GLM (Supplemental Figure \ref{fig:clustering}C and D). The second, cluster 2b, shows stronger up-regulation for \nbCtwob\ genes and the dip in expression at larval L3 (12 hrs), including the majority of both up-regulated classes identified in the life-cycle GLM (\DEisoglmCtwob/\DEisoglmup, \DEisoglmCtwobpercent\%) (Supplemental Figure \ref{fig:clustering}E and F).  Overall, these clustering results support the main conclusions of the differential expression analysis that only a small proportion of \textit{Wolbachia} genes show robust differences in expression across the modENCODE life-cycle time course at the level of the whole organism, and that the majority of dynamically-expressed genes show up-regulation after embryogenesis. However, the greater number of genes identified in cluster 2 relative to those identified by the life-cycle GLM suggests that our differential expression analysis may have only detected a conservative subset of \textit{Wolbachia} genes that show the strongest expression differences across \textit{D. melanogaster} development. 

\subsection*{Dynamically-expressed \textit{Wolbachia} genes are predicted to be involved in stress response and host-microbe interactions}

The \DEisoglm\ \textit{Wolbachia} genes that exhibit dynamic expression across the modENCODE \textit{D. melanogaster} life-cycle time course fall into three broad classes (Figure \ref{fig:heatmap_iso1DE}). The first is a small class of \numberstringnum{\DEisoglmclassone}\ genes that show high relative expression in embryos with down-regulation later in the life cycle. Three of these genes are involved in chaperone function (GroES/WD0308, DnaK/WD0928, and Hsp90/WD1277). The chaperone GroEL/WD0307, which putatively forms a complex with GroES/WD0308, is co-transcribed with GroES/WD0308 and shows similar down-regulation at later stages of the life cycle, but does not pass the significance threshold in the life-cycle GLM (adjusted p-value=0.15). Both GroES/WD0308 and GroEL/WD0307 are in top 15 most abundant transcripts based on average TPM across all stages, confirming that chaperones are among the most highly expressed genes in \textit{Wolbachia} \citep{bennuru_stage-specific_2011,darby_analysis_2012,darby_integrated_2014}. High basal expression of GroEL or other chaperone proteins was suggested to be a compensatory mechanism for the accumulation of slightly deleterious non-synonymous mutations in endosymbionts that arise because of their small populations size and lack of recombination \citep{moran_accelerated_1996,fares_evolution_2002}. The differential expression of \textit{Wolbachia} chaperones during the \textit{D. melanogaster} life cycle that we observe may result from different exposure to external sources of stress or different requirements for protein folding/stability between eggs and larvae versus pupae and adults.

The second class, comprising the majority of up-regulated genes detected (\DEisoglmclasstwo/\DEisoglm), shows increases in relative expression starting with the larval L1 or L2 stages carrying on into adulthood, with decreases at the larval L3 (12 hr) stage and increases at the white pre-pupal 2 and 3 day stages. Genes in this class are mostly unannotated, but include eight genes that code for proteins with membrane or secretion system function (WspB/WD0009, TerC/WD0194, SPFH domain/WD0482, type II secretion/WD0500, HlyD/WD0649, type I secretion/WD0770, VirB3/WD0859, Rhoptry surface protein related/WD1041) and four ANK-containing genes (WD0191, WD0385, WD0438, WD1213). ANK-containing genes from several bacterial species have been shown to be type IV secretion system effector molecules that have diverse effects on eukaryotic cells \citep{caturegli_anka:_2000,sisko_multifunctional_2006,lin_anaplasma_2007,pan_ankyrin_2008,obrien_legionella_2015}. Thus, these results suggest secretion of ANK-containing genes into the host cell may be enriched during early larval and mid-to-late pupal stages of \textit{D. melanogaster} development. Up-regulation of components for secretion systems (type III) has been observed in pupal stages for other arthropod endosymbionts \citep{dale_type_2002}, suggesting that metamorphosis may be a general period that is enriched for up-regulation of secreted symbiont effector proteins involved in host interaction. The presence of a homologue for the \textit{E. coli} transcriptional regulator DksA/WD1094 in this class also provides a potential mechanism to understand the common differential regulation of these genes \citep{paul_dksa_2005,costanzo_ppgpp_2008}. 

A third class of \DEisoglmclassthree\ \textit{Wolbachia} genes show up-regulation primarily in \textit{D. melanogaster} adults, with higher expression in adult males relative to adult females at the same age (see more below). Most of the genes in this class are also unannotated, however three are ANK-containing genes (WD0291, WD0292, WD0438). Our observation of sex-specific expression of ANK-containing genes based on global gene expression profiles of \textit{Wolbachia} in \textit{D. melanogaster} extends results from targeted RT-PCR analysis in \textit{Wolbachia} strains from other insects  \citep{sinkins_wolbachia_2005,duron_variability_2007,klasson_mosaic_2009,papafotiou_regulation_2011,wang_large_2014}, and provides further evidence for their possible involvement in cytoplasmic incompatibility. Finally, we note that our qualitative classification of up-regulated genes in classes 2 and 3 is not mutually exclusive, and the existence of four genes (WD0438, WD1288, WD1289 and WD1290) with sex-biased expression that also show differential expression at larval or pupal stages suggests possible shared regulation of these classes. 

\subsection*{RT-qPCR supports patterns of \textit{Wolbachia} gene expression inferred from whole organism RNA-seq}

We validated the major pattern of \textit{Wolbachia} gene expression dynamics between embryos and adults based on RNA-seq by performing reverse-transcriptase real-time quantitative PCR (RT-qPCR) for a sample of 10 genes in the BDSC ISO1 sub-strain. To assess if our findings were restricted to only one specific host strain, we also performed RT-qPCR for the same genes as well as in a second \textit{D. melanogaster} strain (w1118) that carries a \textit{w}Mel genotype very closely related to the one infecting ISO1 (see Figure \ref{fig:combined_figure}A) \citep{chrostek_wolbachia_2013}.  Three \textit{Wolbachia} genes that were not identified as differentially expressed were chosen as reference genes (WD1043, Wsp/WD1063, petB/WD1071) on the basis of a low fold-change and low coefficient of variation across the RNA-seq time course. One of these (WD1063) is Wsp, the most highly expressed gene in the \textit{w}Mel genome that has been used previously as a control for RT-qPCR-based expression analysis in \textit{Wolbachia} \citep{papafotiou_regulation_2011}. We chose five genes that were up-regulated after embryogenesis (WspB/WD0009, WD0061, WD0830, WD0837, WD1289) and two genes that were down-regulated after embryogenesis (groES/WD0308, Hsp90/WD1277) to validate results of the GLM-based RNA-seq differential expression analysis. We performed RT-qPCR on 16-18 hr old embryos, 1-day post-eclosion males, and 1-day post-eclosion virgin females. We measured relative expression levels normalized to the average expression in embryonic samples of the three stably-expressed genes. 

\begin{figure}[ht] \centering
\includegraphics[width=0.7\textwidth]{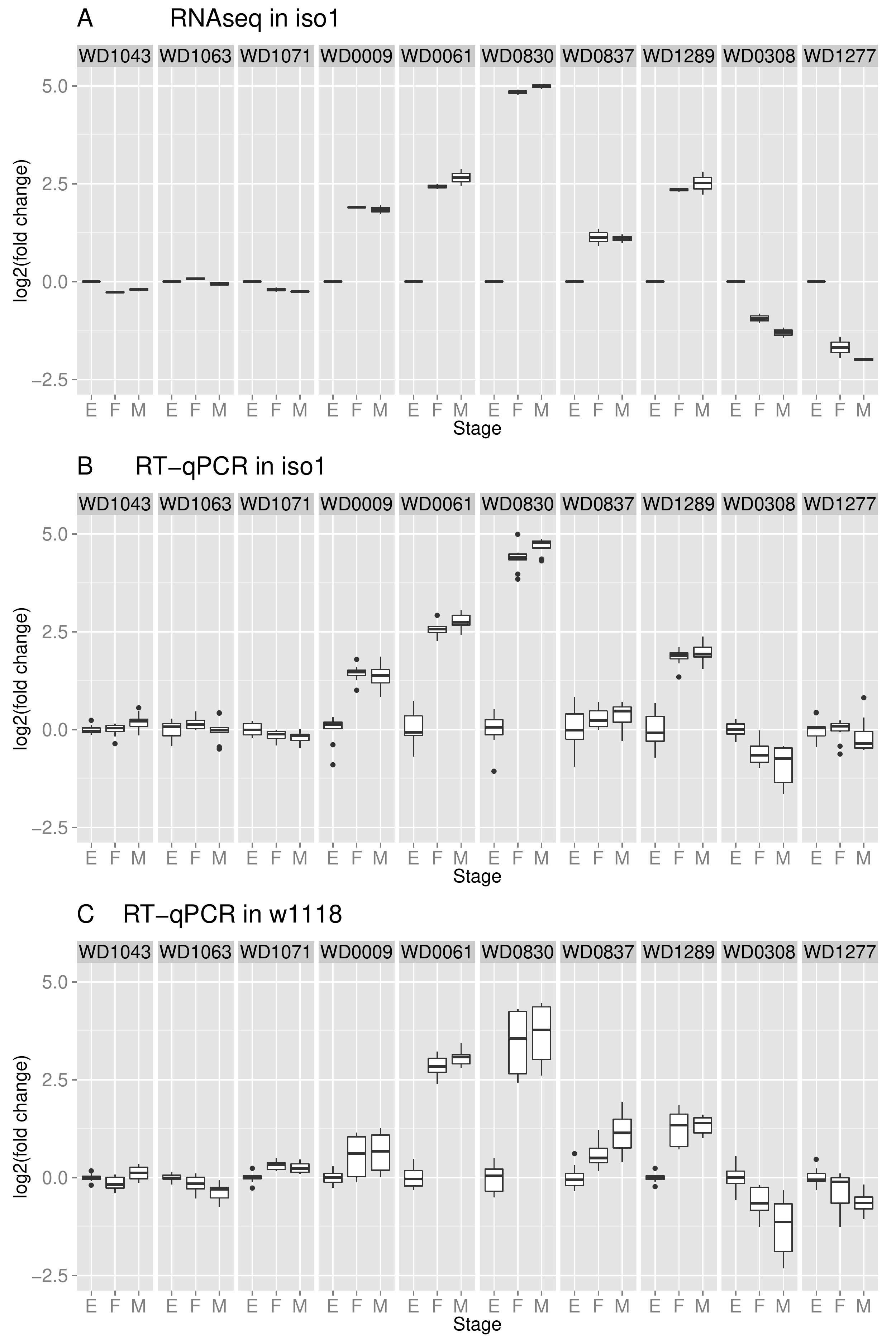}
\caption{{\bf Confirmation of stably- and differentially-expressed genes by RT-qPCR.} A. Relative expression based on RNA-seq in ISO1. B. Relative expression based on RT-qPCR for ISO1. C. Relative expression based on RT-qPCR for w1118. Stages correspond to embryo 16-18 hrs (E), 1-day post-eclosion females (F), and 1-day post-eclosion males (M). RNA-seq expression levels for each gene were based on TPMs and normalized relative to embryonic expression levels for that gene. RT-qPCR expression levels for each gene were normalized using the mean expression of three stably-expressed reference genes (WD1043, WD1063, WD1071) and are calculated relative to embryonic expression levels. Relative expression levels are shown as boxplots with boxes representing the interquartile range (IQR), black lines representing median values, and dots representing samples that lie outside 1.5 x IQR. In A, there is one biological replicate for the embryonic stage and two replicates per stage for males and females. In B and C, for each stage in each genotype, there are ten biological replicates from two independent collections (five replicates from each collection). Genes predicted to be stably-expressed, up-regulated (WspB/WD0009, WD0061, WD0830, WD0837, WD1289) or down-regulated (groES/WD0308, Hsp90/WD1277) between embryos and adults by RNA-seq showed expected patterns by RT-qPCR. Results of GLMs for differences in RT-qPCR expression levels between stages can be found in Supplemental File \ref{itm:supplemental_file_3.tsv}.}
\label{fig:qpcr} 
\end{figure}

As shown in Figure \ref{fig:qpcr}, RT-qPCR in both ISO1 and w1118 broadly confirmed that genes predicted to be stably-expressed based on RNA-seq do not change substantially in relative expression from embryonic to adult stages, whereas both up-regulated and down-regulated genes do, and in the same direction and relative magnitude as predicted by the RNA-seq analysis.  Generalized linear modelling of RT-qPCR expression levels for ISO1 and w1118 separately revealed more genes with significant differences between embryos and adults in w1118 (upregulated: 5/5 validated; down-regulated: 2/2 validated) than in ISO1 (upregulated: 4/5 validated; down-regulated: 1/2 validated) (Supplemental File \ref{itm:supplemental_file_3.tsv}). We also detected slight but significant differences in RT-qPCR expression between embryos and adults for two genes in the ``stable" class in w1118 (Wsp/WD1063, petB/WD1071), which are not observed in ISO1. 

To increase overall power and to separate gene-specific from strain-specific effects, we analysed RT-qPCR data for both strains together in a single GLM (Supplemental File \ref{itm:supplemental_file_3.tsv}). This joint analysis revealed that all seven genes predicted to be up- or down-regulated based on RNA-seq in ISO1 were confirmed as differentially expressed between embryos and adults using RT-qPCR. Differential expression was observed for all genes in both sexes with the exception of Hsp90/WD1277, which only showed significant differences between embryos and males. Genes predicted to be stably-expressed are also confirmed, with the exception of WD1043 which shows a slight, but marginally significant up-regulation in males versus embryos.  Overall, we conclude that RT-qPCR results validate the \textit{Wolbachia} gene expression dynamics inferred from whole-organism RNA-seq in \textit{D. melanogaster}, and that gene expression information from the ISO1 RNA-seq time course can likely be extrapolated to other strains carrying \textit{w}Mel-like \textit{Wolbachia} infections.

\subsection*{\textit{Wolbachia} genes with sex-biased expression show age-dependent effects}

\begin{figure}[ht] \centering
\includegraphics[width=1.0\textwidth]{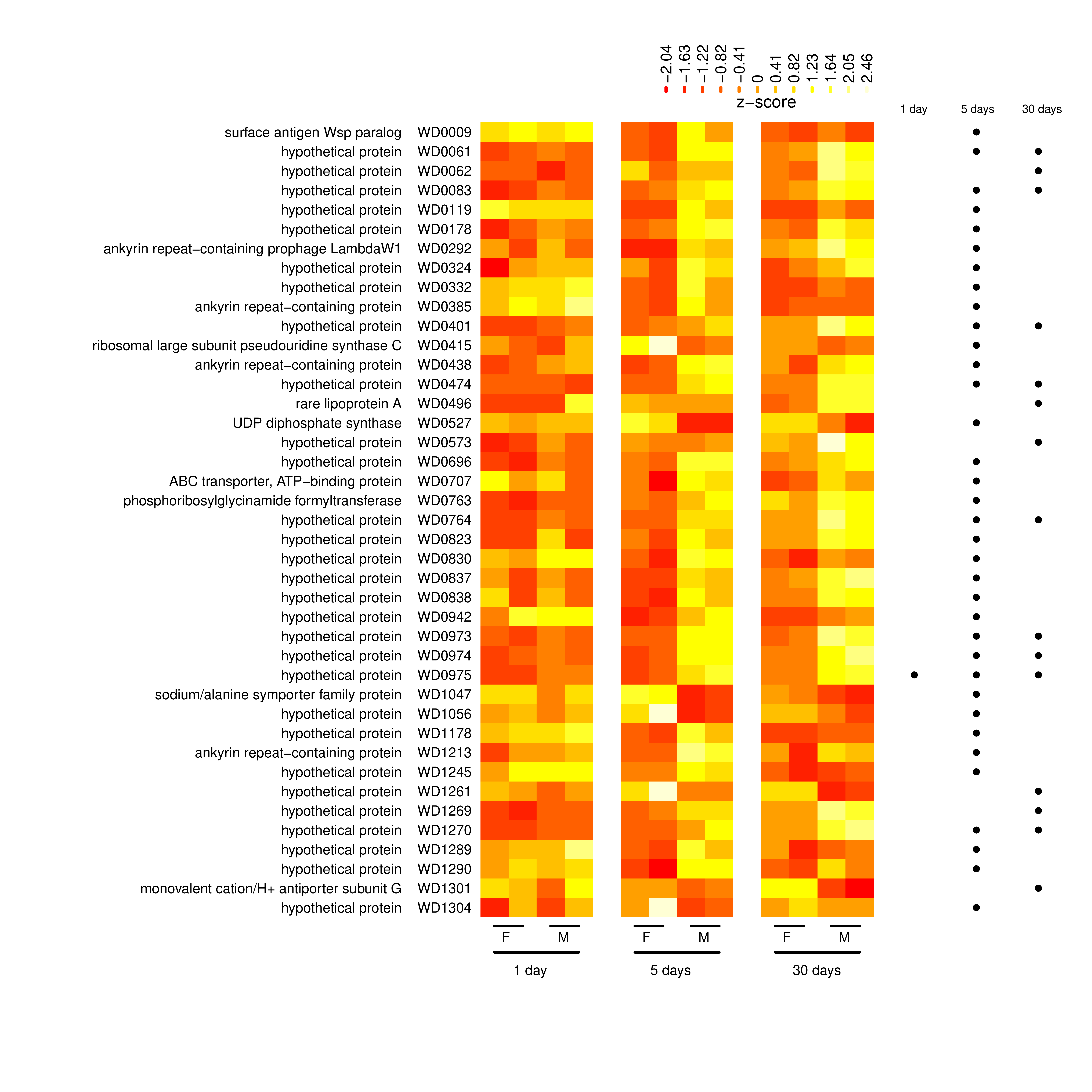}
\caption{{\bf \textit{Wolbachia} genes show age-dependent sex-biased expression.} Row-normalized expression levels are visualized as a heatmap where each row represents a gene (ordered top-to-bottom by its position in the genome), and each cell represents the relative expression level for a particular sample in terms of Z-scores (observed TPM minus row mean TPM, divided by the standard deviation of TPMs for that row). Values higher than row means are represented by yellow, and values lower than row means are represented by red.  Gene names and identifiers are shown on the left. Dots on the right indicate if a gene is differentially expressed between males and females at 1 day, 5 days or 30 days post-eclosion, respectively. All \DEImfall\ genes identified as differentially expressed in pairwise comparisons between males and females at any age in ISO1 are shown here.}
\label{fig:heatmap_isoMFall} 
\end{figure}

As \textit{Wolbachia} is known to cause a variety of sex-specific phenotypes on its hosts \citep{werren_wolbachia:_2008}, including cytoplasmic incompatibility in \textit{D. melanogaster} \citep{hoffmann_partial_1988}, we performed a more detailed analysis of \textit{Wolbachia} expression between males and females at matched ages. For this analysis, we used an exact testing framework \citep{robinson_small-sample_2008} because the GLM-based approach used for the complete life cycle is not the optimal method to use in a pairwise context. We identified a total of \DEImfall\ genes that exhibited greater than 1.5-fold difference at an adjusted p-value cutoff of 0.01 in pairwise tests between male and female samples at either 1, 5 or 30 days post-eclosion, respectively (Figure \ref{fig:heatmap_isoMFall}, Supplemental File \ref{itm:supplemental_file_2.tsv}). Most sex-biased genes in this analysis were identified in one or both of the up-regulated classes in the life-cycle GLM above (\DEImfallGLM/\DEImfall, \percentDEImfall\%), indicating these complementary approaches identify a similar set of \textit{Wolbachia} genes with detectable sex-biased expression in the modENCODE data set. Likewise, sex-biased genes comprise over one-third of differentially-expressed genes identified in the life-cycle GLM (\DEImfallGLM/\DEisoglm, \percentDEImfallGLM\%), suggesting that sex-biased expression is a dominant component of major differences in \textit{Wolbachia} gene expression that can be observed across the \textit{D. melanogaster} life cycle. Neither the GLM nor pairwise analysis revealed sex-biased expression in \textit{D. melanogaster} for homologs of \textit{Wolbachia} genes from \textit{Culex pipiens} (WD0631, WD0632; WD0254, WD0255, WD0508, WD0622, WD0623, WD0626) that have recently been suggested to play a role in cytoplasmic incompatibility \citep{beckmann_detection_2013,pinto_transcriptional_2013}.

Many \textit{Wolbachia} male-biased genes identified in either the life-cycle GLM or male-vs-female pairwise analyses are found consecutively in the genome in small clusters (WD0061--WD0062, WD0291--WD0292, WD0763--WD0764, WD0837--WD0838, WD0973--WD0975, WD1269--WD1270 and \seqsplit{{WD}1288}--\seqsplit{{WD}1290}), suggesting that they might be co-transcribed as operons (see example in Figure \ref{fig:combined_figure}B). Visual inspection confirmed that six out of seven of these clusters (WD0061--WD0062, WD0291--WD0292, WD0837--WD0838, WD0973--WD0975, WD1269--WD1270 and \seqsplit{{WD}1288}--WD1290) are on the same strand and co-transcribed as operons based on contiguous mapping of RNA-seq reads between genes (Supplemental Figure \ref{fig:operons}). The remaining cluster (WD0763--WD0764) corresponds to two divergently-transcribed genes whose co-regulation may represent transcription from a common promoter region. While operon structure is predicted to be common in the \textit{Wolbachia} genome \citep{pertea_operondb:_2009,dehal_microbesonline:_2010,taboada_proopdb:_2012,mao_door_2014}, the detection of sex-biased expression for individual genes in the same operon mutually reinforces that these genes are truly differentially expressed, and suggests common functionality of the genes in these as-yet uncharacterized loci.

The majority of sex-biased genes in the pairwise male-vs-female analyses showed higher expression in males relative to females at matched stages, with only \numberstringnum{\DEImfupFem}\ genes (rluC/WD0415, uppS/WD0527, \seqsplit{sodium/alanine} symporter/WD1047, WD1056, WD1261, cation antiporter subunit G/WD1301, WD1304) showing relatively higher expression in females at one or more time point. Additionally, most genes with sex-biased expression were identified at 5 days post-eclosion (\DEImffive/\DEImfall), many of which maintained sex-biased expression until 30 days post-eclosion. Whole-organism RNA-seq at 5 days post-eclosion correctly predicted the presence (3/3) or lack (13/14) of sex-biased expression differences for 16/17 ANK-containing genes in a \textit{w}Mel strain previously classified by RT-qPCR to have over 1.5-fold difference in expression level between testes and ovaries of 2-day old flies \citep{papafotiou_regulation_2011} (the only exception being that WD0292 shows sex-biased expression in the RNA-seq data at 5-days that is not observed in the RT-qPCR at 2-days). The general lack of sex-biased expression at 1-day post-eclosion inferred from RNA-seq is also supported by our RT-qPCR results (Figure \ref{fig:qpcr}). The five up-regulated genes we tested using RT-qPCR are all sex-biased at 5-days but not 1-day post-eclosion (Figure \ref{fig:heatmap_isoMFall}), none of which show sex-biased expression at 1-day post-eclosion in the RT-qPCR data. 

Our finding that \textit{Wolbachia} genes with sex-biased expression also show age-dependent effects is consistent with a decline in the strength of cytoplasmic incompatibility reported in males at 1-day versus 5-day post eclosion in \textit{D. melanogaster} \citep{reynolds_male_2002,reynolds_effects_2003,yamada_male_2007}. The observed pattern of sex-biased genes being up-regulated in older males is compatible with these \textit{Wolbachia} genes playing a role in attenuating the modification of \textit{D. melanogaster} sperm that leads to embryonic lethality in incompatible crosses \citep{poinsot_mechanism_2003}. Alternatively, if the host is responsible for reducing the effects of \textit{Wolbachia} on the sperm of older males, up-regulation of \textit{Wolbachia} genes in older males could represent a compensatory effect and hence indicate these genes play a role in promoting cytoplasmic incompatibility.

\subsection*{\textit{w}Mel genes with dynamic expression in \textit{D. melanogaster} are conserved in other \textit{Wolbachia} strains}

To identify if candidate genes identified on the basis of differential expression in \textit{D. melanogaster} might interact more broadly with other hosts, we asked whether \textit{w}Mel genes that show stage- and sex-specific expression are conserved in other divergent \textit{Wolbachia} strains.  For this analysis, we used complete \textit{Wolbachia} genome sequences from \textit{w}Ri (an arthropod supergroup A strain from \textit{D. simulans}), \textit{w}Pip-Pel (an arthropod supergroup B strain from \textit{Culex quinquefasciatus}) and \textit{w}Bm (a nematode supergroup D strain from \textit{Brugia malayi}) \citep{klasson_genome_2008,klasson_mosaic_2009,foster_wolbachia_2005}. We identified and clustered homologs in all genomes analyzed, and reconstructed homology groups that included \textit{w}Mel homologs for \DEallwMel\ of \DEall\ genes that show either stage- or sex-specific expression (\numberstringnum{\DEallnowMel}\ dynamically-expressed \textit{w}Mel genes were too small to pass BLAST filtering cutoffs). Only \numberstringnum{\DEallwMelMelanogaster}\ of the \DEallwMel\ dynamically-expressed genes in homology groups (\DEallwMelMelanogasterpercent\%) were restricted to the \textit{w}Mel genome, whereas \DEallwMelArthropod\ genes (\DEallwMelArthropodpercent\%) had homologs in \textit{Wolbachia} genomes from other arthropods, and a further \DEallwMelEcdysozoan\ genes (\DEallwMelEcdysozoanpercent\%) also had homologs in \textit{Wolbachia} genomes from nematodes. The phylogenetic distribution of dynamically-expressed genes does not differ from genome-wide expectations (${\chi}^2=\chisq, p=\chisqpVal, d.f.=2$). These results indicate that the majority of genes identified as dynamically expressed in \textit{D. melanogaster} are core components of the \textit{Wolbachia} gene repertoire and are not unusual in their degree of conservation. Nevertheless, many dynamically-expressed candidate genes are arthropod-specific and few are \textit{D. melanogaster}-specific, as might be expected for a facultative endosymbiont that can switch arthropod hosts by horizontal transfer. Arthropod-specific dynamically-expressed genes include several ANK-containing genes (WD0191, WD0636, WD1213) and membrane/secretion system genes (ABC transporter/WD0455, SPFH domain/WD0482, type II secretion/WD0500, \seqsplit{sodium/alanine} symporter/WD1047, ClpA/WD1237), emphasizing the importance of inter-cellular communication in explaining how \textit{Wolbachia} forms facultative symbioses with its arthropod hosts.

\subsection*{Expression of \textit{Wolbachia} genes previously implicated in host-microbe interaction}

\begin{figure}[ht] \centering
\includegraphics[width=0.7\textwidth]{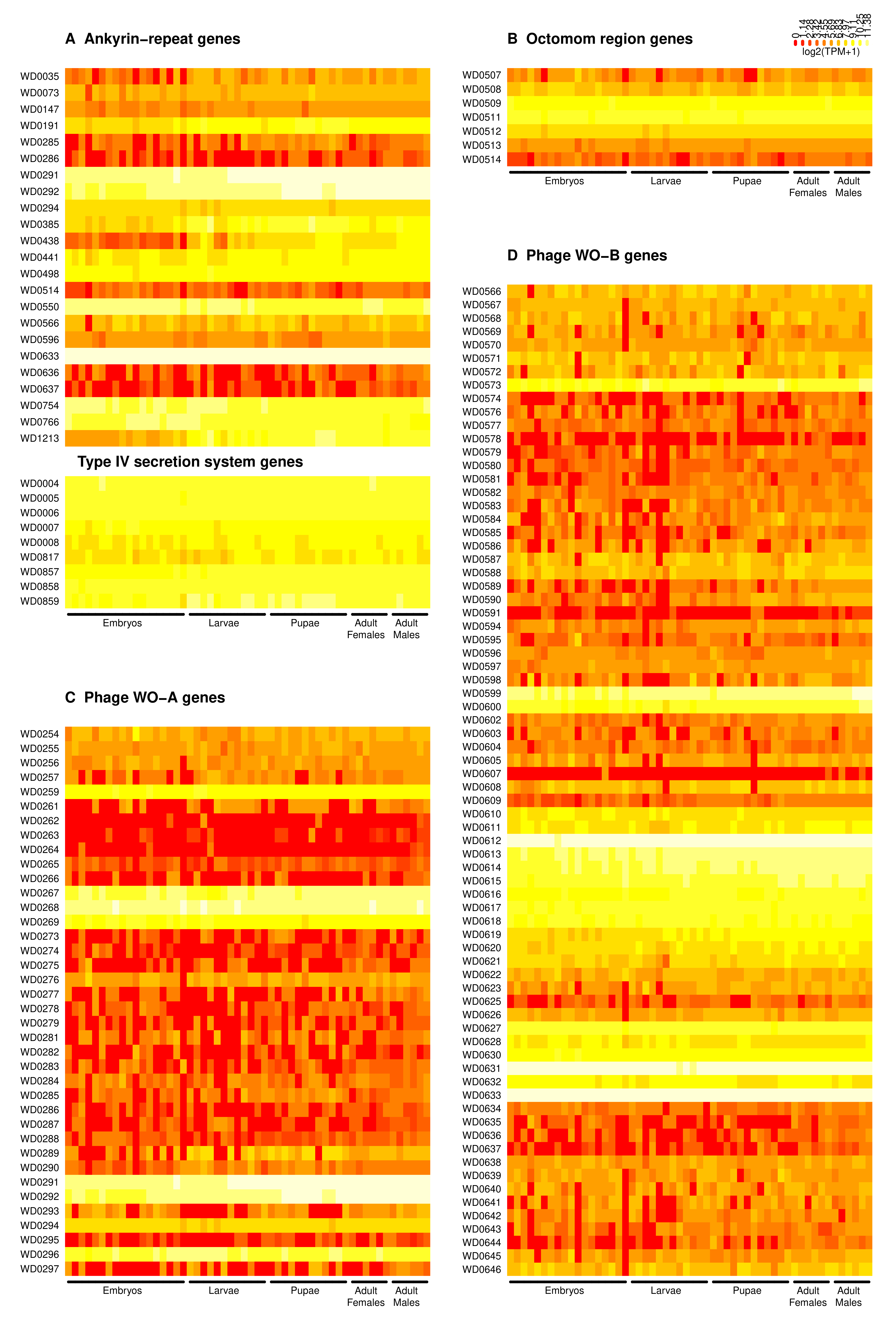}
\caption{{\bf Expression profiles of \textit{Wolbachia} genes previously implicated in host-microbe interactions.} A. ANK-containing and type IV secretion system genes. B. Octomom genes. C. Prophage WO-A genes. D.  Prophage WO-B genes. The 23 ANK-containing genes in panel A are distributed throughout the \textit{Wolbachia} genome. The nine Type IV secretion system genes are found in three different genomic intervals. The Octomom, prophage WO-A and prophage WO-B regions are each from single intervals in the \textit{Wolbachia} genome. The Octomom region only contains seven genes, since WD0510 is not included in the Ensembl annotation for \textit{Wolbachia} \textit{w}Mel. Phage coordinates are from \citet{metcalf_recent_2014}. Expression levels are visualized as a heatmap where each row represents a gene (ordered top-to-bottom by its position in the genome) and each cell represents expression in units of TPM. A pseudocount of one was added to each gene's TPM before transforming to log2 scale. Values with higher levels of expression are represented by yellow, and values with lower levels of expression are represented by red.  All panels are on the same heatmap color scale. Gene names and identifiers are shown on the left. All stages including those that lack biological replicates in the modENCODE time course are shown here.}
\label{fig:candidate_genes} 
\end{figure}

In addition to identifying genes that are candidates for mediating host-microbe interaction on the basis of their stage- or sex-specific differential expression, we also investigated expression levels of \textit{Wolbachia} genes previously suggested to be candidates for mediating interaction with \textit{D. melanogaster}. The most-widely hypothesized set of candidates for host-microbe interaction are the 23 ANK-containing genes that are possible type IV secretion system effectors in \textit{Wolbachia} \citep{wu_phylogenomics_2004,iturbe-ormaetxe_distribution_2005,papafotiou_regulation_2011,siozios_diversity_2013}, which the modENCODE data show are expressed at widely different levels in \textit{D. melanogaster} (Figure \ref{fig:candidate_genes}A). The five most weakly-expressed ANK-containing genes (WD0285, WD0286, WD0514, WD0636, WD0637) are found in the Octomom and prophage regions, and are the same five genes which \citet{papafotiou_regulation_2011} found were expressed too weakly to obtain reliable RT-qPCR data in adult gonads. Thirteen ANK-containing genes are highly expressed (WD0191, WD0291, WD0292, WD0294, WD0385, WD0438, WD0441, WD0498, WD0550, WD0633, WD0754, WD0766, WD1213), which include the majority of ANK genes identified as differentially expressed in this study  or by \citet{papafotiou_regulation_2011} (WD0191, WD0291, WD0292, WD0294, WD0385, WD0438, WD0550, WD0636, WD1213). The nine genes that make a complete type IV secretion system in the \textit{w}Mel are all highly expressed in all stages, including the virB8 paralog (WD0817) which is not a part of the two genomic clusters that contain the remaining eight type IV secretion system genes. These results support a model where a functionally competent type IV \textit{Wolbachia} secretion system is expressed throughout the \textit{D. melanogaster} life cycle, with both constitutive and regulated secretion of subsets of ANK-containing effectors. 

The Octomom region is part of the accessory genome of \textit{Wolbachia} \citep{iturbe-ormaetxe_distribution_2005,chrostek_wolbachia_2013} and contains eight genes whose copy number controls \textit{Wolbachia} growth and pathogenesis \citep{chrostek_mutualism_2015}. In the \textit{w}Mel variant where Octomom is present in one copy, all genes in this region (WD0507--WD0514) are expressed at relatively constant levels across the life cycle (Figure \ref{fig:candidate_genes}B). None of these genes show a significant change in gene expression during host development in the life-cycle GLM. However, different Octomom genes do vary considerably in their expression levels relative to each other, with the most highly expressed genes being found in the middle of this interval (WD0509--WD0512). Given that two genes with possible regulatory (the helix-turn-helix containing gene WD0508) or effector (the ANK-containing gene WD0514) function are relatively weakly expressed in the non-pathogenic \textit{w}Mel variant, it is possible that over-expression of one or both of these genes may be responsible for the pathogenic phenotype when Octomom is amplified in \textit{w}MelPop \citep{chrostek_mutualism_2015}.

Unlike obligate endosymbionts with streamlined genomes, prophages are often present in \textit{Wolbachia} from arthropod hosts and have been suggested to directly or indirectly influence \textit{Wolbachia}-host interactions \citep{kent_phage_2010,metcalf_recent_2014}. Two major prophage regions are present in the \textit{w}Mel genome --- called WO-A and WO-B --- both of which have undergone degeneration and rearrangement since insertion \citep{wu_phylogenomics_2004,kent_evolutionary_2011}. There is no clear evidence that the WO-A and WO-B prophages from \textit{w}Mel can enter a lytic phase as they can in other arthropods \citep{masui_bacteriophage_2001,fujii_isolation_2004,bordenstein_tripartite_2006,sanogo_wo_2006}, however phage-like particles have been reported in extracts of \textit{D. melanogaster} strains infected with \textit{w}Mel \citep{gavotte_diversity_2004}. Consistent with previous results from \textit{w}MelPop-CLA \citep{darby_integrated_2014} and prophages in \textit{Salmonella enterica} \citep{perkins_strand-specific_2009}, expression levels of most genes in the WO-A and WO-B prophage regions are typically very low across the entire \textit{D. melanogaster} life cycle (Figure \ref{fig:candidate_genes}C and D), and define the largest segments of the \textit{w}Mel genome with consecutive lowly-expressed genes Supplemental Figure \ref{fig:genome_heatmap}). The most conspicuous exception to this pattern is the 21-gene interval in WO-B (WD0611--WD0634) that contains genes laterally-transferred between \textit{Wolbachia} and the \textit{Rickettsia} endosymbiont of the tick \textit{Ixodes scapularis} (WD0612--WD0621) \citep{ishmael_extensive_2009,gillespie_rickettsia_2012}, a region not typically present in WO prophage from other \textit{Wolbachia} strains \citep{kent_evolutionary_2011}. In addition, two very abundant anti-sense ncRNA transcripts can be detected overlapping the major capsid genes of both WO-A (WD0274) and WO-B (WD0604) (Supplemental Figure \ref{fig:ncRNAs}), which may play a role in the regulation of prophage genes.

Most prophage-encoded structural genes are expressed at low levels, with only genes in the tail (WD0567--WD0575) and base-plate (WD0638--WD0644) regions of WO-B being expressed at appreciable levels. Likewise, most non-structural prophage-encoded genes previously suggested to be candidates for host interaction (VrlC.2/WD0579, VrlC.1/WD0580, Patatin/WD0565, DNA methylases WD0263 and WD0594) \citep{kent_phage_2010} are expressed at low levels.  Intriguingly, each prophage region contains a highly-expressed operon (WD0267--WD0269 in WO-A and WD0599--WD0600 in WO-B) that encodes homologs of the \textit{E. coli} RelE toxin (WD0269 and WD0600) (Figure \ref{fig:candidate_genes}C and D). RelE is a stress-inducible cytotoxic translational repressor that is counter-acted by the antitoxin RelB, a small protein which gene is co-transcribed in the same operon as RelE \citep{christensen_rele_2001,pedersen_bacterial_2003,yamaguchi_toxin-antitoxin_2011}. The genes adjacent to the RelE homologues in WO-A and WO-B (WD0267, WD0268 and WD0599) are also co-transcribed and encode small peptides, and thus could be acting as antitoxins. In fact, WD0268--WD0269 and WD0599--WD0600 have been computationally predicted to be toxin-antitoxin (TA) pairs \citep{shao_tadb:_2011}, and TA pairs have been previously reported in other cryptic prophages \citep{van_melderen_bacterial_2009}. RelE-containing gene clusters are also found in similar positions (between the terminase and portal genes that form the phage head) in divergent prophages from \textit{D. simulans} (WOriA) and \textit{N. vitripinnis} (WOVitA2) \citep{kent_evolutionary_2011}, further indicating they may play some conserved functional role such as stabilizing the \textit{Wolbachia} prophage genomic regions by preventing large-scale deletions \citep{van_melderen_bacterial_2009}.  Low expression levels of phage structural genes, together with highly-expressed putative TA pairs suggests that prophage in \textit{w}Mel are maintained the lysogenic state by self-preservation.

\clearpage

\section*{Conclusions}

We have shown that the ISO1 reference strain used by the modENCODE project is infected with \textit{Wolbachia}, and used this fortuitous observation to study the global expression dynamics of a facultative endosymbiont over the life cycle of the model insect species, \textit{D. melanogaster}. Our work represents the most comprehensive gene expression profiling  to date of an endosymbiotic bacteria in its native host context. We establish that most \textit{Wolbachia} genes are expressed in all \textit{D. melanogaster} life-cycle stages, but that major changes in expression levels of \textit{Wolbachia} genes are rare when studied simultaneously across all \textit{D. melanogaster} tissues. Additional work is needed to rule out lack of power to detect real differences due to the low number of replicates and averaging of tissue-specific expression in our study, but global stability in \textit{Wolbachia} expression is mechanistically consistent with the limited number of regulatory genes encoded in the \textit{w}Mel genome \citep{wu_phylogenomics_2004}. Global stability of \textit{Wolbachia} gene expression across the \textit{D. melanogaster} life-cycle may be an adaptive response that simply reflects the stable environment of an intracellular endosymbiont, or that suggests \textit{Wolbachia} co-exists in non-obligatory symbioses largely in a stable ``stealth mode" that reduces the chances of being detected by the host as a pathogen.  

Nevertheless, a set of \DEall\ \textit{Wolbachia} \textit{w}Mel genes that show robust stage- or sex-specific differential expression can be identified at the whole-fly level, many of which share common expression dynamics and therefore may be co-regulated. These genes provide many new candidate genes for understanding, and possibly manipulating, the genetic basis of how \textit{Wolbachia} interacts with arthropod hosts. Furthermore, differences in expression levels among \textit{Wolbachia} genes previously implicated to mediate host-symbiont interaction can provide insight into the likelihood and mechanistic basis of existing candidates. Importantly, we also provide the first detailed insight into the developmental dynamics of \textit{Wolbachia} gene expression in an insect host, which suggest larval and pupal stages (where \textit{Wolbachia} have been detected cytologically \citep{clark_widespread_2005}) merit further study to understand how \textit{Wolbachia} manipulates host biology to maintain persistent infections and affect transmission. 

Future studies can leverage our finding that the modENCODE total RNA-seq dataset contains a nearly-complete \textit{Wolbachia} transcriptome to functionally annotate the transcriptional landscape of the \textit{Wolbachia} genome. Currently, only protein-coding regions and a small number of ncRNAs are included in the \textit{w}Mel genome annotation, and recent work has identified a handful of additional \textit{Wolbachia} ncRNAs \citep{mayoral_wolbachia_2014,woolfit_discovery_2015}. The strand-specific total RNA-seq data from the modENCODE can now be used to generate high-quality transcript models to annotate 5' and 3' untranslated regions, operons and identify new ncRNA genes in \textit{Wolbachia} (see examples in Supplemental Figures \ref{fig:operons} and \ref{fig:ncRNAs}). The possibility of a more comprehensive annotation of ncRNAs in \textit{Wolbachia} is particularly exciting given recent work suggesting ncRNAs provide an important layer of post-transcriptional regulation to modulate protein expression levels in the \textit{Buchnera} endosymbiont of aphids \citep{hansen_widespread_2014}. Together with other recently published transcriptomic data \citep{darby_integrated_2014,mayoral_wolbachia_2014,woolfit_discovery_2015}, the necessary materials are now available to systematically reannotate the \textit{Wolbachia} \textit{w}Mel genome in order to support basic and applied research on this important model organism.

\clearpage

\section*{Methods}
\label{sec:Methods}

\subsection*{\textit{D. melanogaster} strains and husbandry}
Sub-strains of the \textit{D. melanogaster} ISO1 strain originally described in \citet{brizuela_genetic_1994} were obtained from several sources: (i) the Bloomington \textit{Drosophila} Stock Center (BDSC ISO1, stock \#2057); (ii) Jim Kennison (National Institute of Child Health and Human Development); (ii) Todd Laverty (Howard Hughes Medical Institute Janelia Farms); and (iv) Sue Celniker (Lawrence Berkeley National Laboratory). The DrosDel w1118 isogenic line carrying \textit{Wolbachia} variant \textit{w}Mel (hereafter ``w1118") is described in \citet{chrostek_wolbachia_2013}. \textit{D. melanogaster} lines were maintained on standard cornmeal diet at a constant temperature of 25\si{\degreeCelsius}. 

We generated versions of all ISO1 sub-strains cured of any potential \textit{Wolbachia} infection by treating with tetracycline for two generations. Adults were allowed to lay eggs for 5 days on Formula 4-24 food (Carolina, cat \#173210) mixed with equal part water containing \SI{0.25}{\micro\gram}/\SI{}{\milli\liter} tetracycline. Offspring from the first generation were collected on day 10 and transferred to new food containing \SI{0.25}{\micro\gram}/\SI{}{\milli\liter} tetracycline and allowed to lay eggs for 5 days. Offspring from the second generation were collected on day 10 and transferred onto standard cornmeal-agar food to establish \textit{Wolbachia}-free stocks.

\subsection*{\textit{Wolbachia} infection status}

DNA for PCR screening of \textit{Wolbachia} infection status was prepared from single flies by placing individual males in a standard fly squish buffer (\SI{50}{\micro\liter} of 1M Tris pH 8.0, 0.5M EDTA, 5M NaCl) plus \SI{1}{\micro\liter} of \SI{10}{\milli\gram/\milli\litre} Proteinase K. Flies were then placed in a thermocycler at 37\si{\degreeCelsius} for 30 minutes, 95\si{\degreeCelsius} for 2 minutes followed by a 4\si{\degreeCelsius} hold. \SI{4}{\micro\liter} of fly squish product was used in a \SI{50}{\micro\liter} PCR. The presence of \textit{Wolbachia} was confirmed by PCR using two sets of primers: (i) Wolbachia\_F2 (5{'} TGGCTCACATAGATGCTGGT 3{'}) and Wolbachia\_R2 (5{'} GTCCCATTTCTCACGCATTT 3{'}); and (ii) Wolbachia\_F3 (5{'} ATCCTGCAAATTGGCGTACT 3{'}) and Wolbachia\_R3 (5{'} ATAACGCACACCTGGCAAAT 3{'}). To ensure DNA preparation was sufficient for PCR amplification, control primers were used from the \textit{D. melanogaster} genome: rDNA-F (5{'} AAACTAGGATTAGATACCCTATTAT 3{'}) and rDNA-R (5{'} AAGAGCGACGGGCGATGTGT 3{'}). PCR was performed with Kappa HiFi polymerase (KAPA Biosystems, KK2502) using the following reaction conditions: 30 cycles of 95\si{\degreeCelsius} for 20 seconds, 60\si{\degreeCelsius} for 15 seconds, 72\si{\degreeCelsius} for 90 seconds.

\subsection*{Genome sequencing and data analysis}

Genomic DNA for the BDSC ISO1 strain was prepared from 10 starved, adult males using the Qiagen DNeasy Blood and Tissue Kit (Qiagen, 69504). \SI{1}{\micro\gram} of DNA from each was fragmented using a Covaris S220 sonicator (Covaris Inc.) to 250 base pair (bp) fragments by adjusting the treatment time to 85 seconds. Following manufacturer's directions, short fragment libraries were made using the KAPA Library Preparation Kits (KAPA Biosystems, KK8201) and Bioo Scientific NEXTflex DNA Barcodes (Bioo Scientific, 514104). The resulting libraries were purified using Agencourt AMPure XP system (Beckman Coulter, A63880), then quantified using a Bioanalyzer (Agilent Technologies) and a Qubit Fluorometer (Life Technologies).  Libraries were pooled with other strains, re-quantified and run for 100 cycles in paired-end high output mode over multiple lanes on an Illumina HiSeq 2000 instrument using HiSeq Control Software v1.5.15.1 and Real-Time Analysis v1.13.48.0. CASAVA v1.8.2 was run to demultiplex reads and generate fastq files. Raw fastq reads were submitted to ENA as experiment ERX645969.

DNA-seq fastq sequences from ERX645969 were downloaded and mapped against a ``holo-genome" consisting of the Release 5 version of the \textit{D. melanogaster} genome (Ensembl Genomes Release 24, Drosophila\_melanogaster.BDGP5.24.dna.toplevel.fa) and the \textit{Wolbachia} \textit{w}Mel reference genome (Ensembl Genomes Release 24, Wolbachia\_endosymbiont\_of\_drosophila\_melanogaster.GCA\_000008025.1.24) \citep{cunningham_ensembl_2015,kersey_ensembl_2014}. Holo-genome reference mapping was performed using bwa mem v0.7.5a \citep{li_aligning_2013} with default parameters in paired-end mode. Mapped reads for all runs from the same sample were merged, sorted and converted to BAM format using samtools v0.1.19 \citep{li_sequence_2009}.  BAM files were then used to create BCF and fastq consensus sequence files using samtools mpileup v0.1.19 (options -d 100000). Fastq consensus sequence files were converted to fasta using seqtk v1.0-r76-dirty (https://github.com/lh3/seqtk) and concatenated with consensus sequences of \textit{Wolbachia} type strains from \citet{chrostek_wolbachia_2013}. Maximum-likelihood phylogenetic analysis on resulting multiple alignments was performed using raxmlHPC-PTHREADS v8.1.16 (options -T 6 -f a -x 12345 -p 12345 -N 100 -m GTRGAMMA) \citep{stamatakis_raxml_2014}. 

\subsection*{RNA-seq data analysis}

We analysed total RNA-seq libraries from the modENCODE developmental time course, which samples 30 time points from the BDSC ISO1 substrain across the \textit{D. melanogaster} life cycle including embryos, larvae, pupae, adult males, and adult females \citep{graveley_developmental_2011,brown_diversity_2014,duff_genome-wide_2015}. Total RNA-seq data from this project is 100 bp read length, paired-end and stranded, with two biological replicates available for 24 of the 30 time points. All non-adult samples are from mixed sex organisms in unknown ratios; adult female samples are from mated and virgin flies in unknown ratios. 

Total RNA-seq fastq sequences from SRP001696 were downloaded and mapped against the holo-genome described above in paired-end mode using bwa mem v0.7.5a with default parameters. Accession numbers for individual samples are given in Supplemental File \ref{itm:supplemental_file_1.tsv}. Resulting mapped reads were sorted and converted to BAM format using samtools v0.1.19. Counts for both forward and reverse reads together were used to summarize numbers of reads mapping to the \textit{Wolbachia} and \textit{D. melanogaster} genomes. Forward reads from each read-pair (which are anti-sense orientation in the Illumina TruSeq Stranded Total RNA kit used) were converted to the opposite strand and combined with reverse reads to generate wiggle plots of strand-specific RNA-seq coverage. 

Sorted BAM files were used to count reads overlapping protein-coding genes on the sense orientation by one or more base pair using BEDtools v2.22.0 \citep{quinlan_bedtools:_2010} with the Ensembl Genomes Release 24 version of the \textit{Wolbachia} genome annotation (Wolbachia\_endosymbiont\_of\_drosophila\_melanogaster.GCA\_000008025.1.24.gtf). Since current gene models in \textit{Wolbachia} correspond only to coding regions and not full-length transcripts, we chose a read counting strategy that allowed RNA-seq reads to extend beyond currently-annotated gene model limits. Only counts for the reverse read from each read-pair (the sense orientation in the Illumina TruSeq Stranded Total RNA kit used) were used for differential expression analysis and clustering. 

We performed differential expression analysis using edgeR v3.6.8 \citep{robinson_edger:_2010} using P-values adjusted with the \citet{benjamini_controlling_1995} method to correct for multiple testing. Read counts were normalized using the trimmed mean of M-values method \citep{robinson_scaling_2010} and models were fit using tagwise dispersion \citep{robinson_moderated_2007}. To identify \textit{Wolbachia} genes that change in any stage across the life cycle, we performed a single analysis using an ANOVA-like GLM approach \citep{mccarthy_differential_2012} using all stages of the ISO1 developmental time course that had replicates (24 time points) with an adjusted p-value cutoff of 0.05. To identify \textit{Wolbachia} genes that change between  pairs of samples we used an exact test approach \citep{robinson_small-sample_2008} with adjusted p-value 0.01 and two-fold change cutoffs. 

Probabilistic clustering on all samples from the ISO1 time course (both with and without biological replicates) was performed with MBcluster.seq v.1.0 \citep{si_model-based_2014} using the Poisson model with two clusters and the expectation maximization method. Because MBcluster.seq is a probabilistic method, we performed 1,000 runs of the clustering analysis. We matched cluster identifiers from different runs using the fact that the majority of genes are stably-expressed across the life-cycle, and defined the cluster with the majority of genes as ``cluster 1" and the remaining genes as ``cluster 2". Genes was assigned to cluster 2 were further classified into sub-clusters 2a and 2b on the basis of the number of runs in which a gene was assigned to cluster 2.

Within-sample normalized read counts in units of transcripts per million (TPM) \citep{li_rsem:_2011} were also used to generate between-sample correlation and gene-by-sample heatmaps. Effective gene length in TPM calculations was set to be $gene\_length + read\_length - 1$ to account for reads that extend beyond annotated gene models. Differential expression and clustering analyses and visualization were performed in R v3.1.1 \citep{r_development_core_team_r:_2012}.

\subsection*{Reverse Transcription Real-Time Quantitative PCR}

RNA for reverse transcription real-time quantitative PCR (RT-qPCR)  was obtained from embryos, adult males and adult virgin females for the BDSC ISO1 and w1118 strains. Two independent collections, each with five biological replicates per stage, were performed for each \textit{D. melanogaster} line. For embryo collection, flies laid eggs for two hours on agar plates supplemented with 1:1 yeast/water paste. After 16 hours at 25\si{\degreeCelsius} the embryos were collected, treated with 2\% sodium hypochlorite, and washed with sterile water before RNA extraction. 500 embryos were used per sample. For adult collection, males and females were separated immediately after eclosion and maintained on standard diet for 24 hours before RNA extraction. Ten adult flies were used per sample. Samples were homogenized with a plastic pestle in \SI{1}{\milli\liter} of Trizol Reagent (Ambion, 15596-018). RNA was extracted according to manufacturer's protocol and resuspended in \SI{50}{\micro\liter} of DEPC-treated water (Ambion, AM9915G). RNA concentrations were determined using NanoDrop ND-1000 Spectrophotometer. cDNA was prepared from \SI{4}{\micro\gram} of total RNA using random primers (Promega, C1181) and M-MLV reverse transcriptase (Promega, M1705). Primers were allowed to bind to the template RNA at 70\si{\degreeCelsius} for 5 minutes and the reaction proceeded at 25\si{\degreeCelsius} for 10 minutes, 37\si{\degreeCelsius} for 60 minutes, and 85\si{\degreeCelsius} for 10 minutes.

RT-qPCR reactions were carried out in CFX384 Real-Time PCR Detection System (BioRad). Reactions were carried in 384-well plates (BioRad, HSP3805) using iTaq Universal SYBR Green Supermix (Bio Rad, 172-5125), \SI{0.15}{\micro\molar} of each primer and \SI{5}{\micro\litre} of cDNA diluted 1:50 in water. Each complete, independent collection of each \textit{D. melanogaster} line was analysed in one plate. Each plate contained two technical replicates of every sample for each set of primers.  Sequences of the primers used for RT-qPCR can be found in Supplemental File \ref{itm:supplemental_file_4.tsv}. Amplification conditions were set up as follows: 50\si{\degreeCelsius} for 2 minutes, 95\si{\degreeCelsius} for 10 minutes, followed by 40 cycles of 95\si{\degreeCelsius} for 30 seconds, 57\si{\degreeCelsius} for 1 minute and  72\si{\degreeCelsius} for 30 seconds. Melting curves were analyzed to confirm specificity of amplified products and Ct values were obtained using Bio-Rad CFX Manager default threshold settings. Relative transcript expression levels were calculated by the method of \citet{pfaffl_new_2001}. Gene expression was normalized using as reference genes the three stably-expressed \textit{Wolbachia} genes WD1043, WD1063, and WD1071, which were selected because they exhibit low fold-change and low coefficient of variation across the ISO1 life-cycle time course \citep{de_jonge_evidence_2007}. Expression values were calculated relative to embryonic expression levels.

Relative gene expression values were analyzed in R v3.1.1 \citep{r_development_core_team_r:_2012} by fitting a linear mixed-effect model to the data of each gene using the lmer package (v2.0-20), comparing the effect of stage (embryo, adult male, and adult female) with a Tukey's all-pair comparison using the glht package (v1.3-9). The data of the two genotypes were analysed separately and together. For the linear mixed-model, the stage and genotype (in the joint analysis) were considered fixed effects, while independent collection was considered a random effect. No correction was applied to p-values to account for multiple testing, and thus $alpha$-levels for significance were set at 0.001 based on Bonferroni correction. 

\subsection*{Functional and comparative annotation of \textit{Wolbachia} genes}
We generated functional annotations for \textit{Wolbachia} genes using three sources: (i) by querying \textit{w}Mel open reading frames against the Genbank nucleotide (nt) database (April 2012, 15,938,872 sequences; 40,783,330,152 letters) using TBLASTN v2.2.25+ \citep{altschul_gapped_1997}; (ii) by querying \textit{w}Mel open reading frames against the Pfam-A.hmm database (v26.0) \citep{finn_pfam:_2014} using hmmscan v3.0 with default options (http://hmmer.org); and (iii) by using the original functional annotations generated by TIGR. CodonW v1.4.4 was used to generate estimates of and GC bias and codon bias in terms of effective number of codons ($N_{c}$) (http://codonw.sourceforge.net).

We identified homologs of \textit{w}Mel genes by conducting an all-vs-all search of genes from the following complete \textit{Wolbachia} genomes using BLASTP 2.2.27+ with default options: wRi (supergroup A strain from \textit{D. simulans}, NC\_012416), \textit{w}Pip-Pel (supergroup B strain from \textit{Culex quinquefasciatus}, NC\_010981), and \textit{w}Bm (supergroup D strain from \textit{Brugia malayi}, NC\_006833). The best hit to a gene in genome A was defined as the gene in genome B that had the highest bit score. Homology groups were defined such that all members of a group had to have reciprocal best hits with all other members of the group (complete linkage), which permits paralogs to be included in a group. Clustering is dependent on the order in which best hits are considered, and thus all best hits were sorted by highest bit score before clustering so that the strongest hits were considered first. 

\clearpage


\section*{List of Abbreviations Used}
\label{List of Abbreviations Used}

ISO1, isogenic strain 1; ANK, ankyrin repeat domain; modENCODE, model organism encyclopedia of DNA elements; \textit{w}Mel, \textit{Wolbachia} from \textit{D. melanogaster}; \textit{w}MelPop, Popcorn strain of \textit{Wolbachia} from \textit{D. melanogaster}; \textit{w}MelCS, \textit{Wolbachia} from Canton S strain of \textit{D. melanogaster}; RNA-seq, RNA sequencing; BDSC, Bloomington \textit{Drosophila} Stock Center; BDGP, Berkeley \textit{Drosophila} Genome Project; L1, first larval instar; L2, second larval instar; L3, third larval instar; TPM, transcripts per million; ANOVA, analysis of variance; GLM, generalized linear model; RT-qPCR, reverse-transcriptase real-time quantitative PCR; ncRNA, non-coding RNA.

\section*{Competing Interests}
\label{Competing Interests}

The authors declare that they have no competing interests.

\section*{Authors' Contributions}
\label{Authors' Contributions}

FG carried the RNA-seq analyses, participated in the design of the study, and helped draft the manuscript. CRC performed RT-qPCR experiments. DEM carried out the \textit{Wolbachia} infection status and genome sequencing experiments. DWR and ILN generated functional and comparative annotations. RSH participated in the coordination of the study. LT participated in the design and coordination of the study, carried out RT-qPCR analyses, and helped draft the manuscript. CMB conceived of the study, participated in its design and coordination, performed phylogenomic analyses, contributed to the RNA-seq analyses, and drafted the manuscript. All authors read and approved the final manuscript. 

\section*{Description of Supporting Data Files}
\label{Description of Supporting Data Files}

The following additional data are available with the online version of this paper. Additional data file 1 is a table summarizing the numbers of mapped reads and expressed genes for each RNA-seq sample. Additional data file 2 is a table summarizing reads counts, expression levels, results of differential expression an functional/comparative annotations of \textit{Wolbachia} genes. Additional data file 3 is a table summarizing the results of RT-qPCR analyses. Additional data file 4 is a table of PCR primers used for the RT-qPCR analyses.

\section*{Acknowledgements}
\label{Acknowledgements}

We thank Jim Kennison, Todd Laverty, and Sue Celniker for providing stocks; Jim Kennison, Gerry Rubin, Todd Laverty and Roger Hoskins for information on the provenance of ISO1 sub-strains; Brent Graveley, Peter Cherbas and Robert Eisman for information about how modENCODE RNA-seq samples were generated; Sarah Bordenstein and Seth Bordenstein for information about prophage coordinates; Raquel Linheiro for R programming guidance; Github for providing free private repositories that enabled this collaboration; and Douda Bensasson, Sarah Bordenstein, Seth Bordenstein, Tim Karr and members of the Bergman Lab for comments on the manuscript. We are especially grateful to Roger Hoskins for his suggestion to look in modENCODE RNA-seq data to confirm our observation that the BDSC sub-strain of ISO1 was infected with \textit{Wolbachia}, which led to the results presented here. This work was supported by a National Environment Research Council Ph.D. studentship (FG), Funda\c{c}\~{a}o para a Ci\^{e}ncia e a Tecnologia post-doctoral fellowship SFRH/BPD/73420/2010 (CRC), an award from the Stowers Institute for Medical Research (RSH), National Science Foundation grant NSF-IOS-1456545 (IGLN), Human Frontier Science Program grant RGY0093/2012 (CMB), and Biotechnology and Biological Sciences Research Council grant BB/L002817/1 (CMB).  

\clearpage

\section*{}
\label{References}
\bibliographystyle{final}
{\footnotesize \linespread{1}
\bibliography{../../../bibtex/bergmanlab.bib}}

\begin{thebibliography}{105}
\newcommand{\enquote}[1]{``#1''}
\providecommand{\natexlab}[1]{#1}
\expandafter\ifx\csname urlstyle\endcsname\relax
  \providecommand{\doi}[1]{doi:\discretionary{}{}{}#1}\else
  \providecommand{\doi}{doi:\discretionary{}{}{}\begingroup
  \urlstyle{rm}\Url}\fi

\bibitem[{Adams \emph{et~al.}(2000)Adams, Celniker
  \emph{et~al.}}]{adams_genome_2000}
Adams, M.D., Celniker, S.E., Holt, R.A., Evans, C.A., Gocayne, J.D.,
  Amanatides, P.G., Scherer, S.E., Li, P.W., Hoskins, R.A., Galle, R.F.,
  George, R.A., Lewis, S.E., Richards, S., Ashburner, M., Henderson, S.N.,
  Sutton, G.G., Wortman, J.R., Yandell, M.D., Zhang, Q., Chen, L.X.,
  \emph{et~al.} (2000).
\newblock \enquote{The genome sequence of {Drosophila} melanogaster.}
\newblock \emph{Science}, \textbf{287}(5461): 2185--2195.

\bibitem[{Altschul \emph{et~al.}(1997)Altschul, Madden
  \emph{et~al.}}]{altschul_gapped_1997}
Altschul, S.F., Madden, T.L., Schäffer, A.A., Zhang, J., Zhang, Z., Miller,
  W., and Lipman, D.J. (1997).
\newblock \enquote{Gapped {BLAST} and {PSI}-{BLAST}: a new generation of
  protein database search programs.}
\newblock \emph{Nucleic Acids Research}, \textbf{25}(17): 3389--3402.

\bibitem[{Andersson and Sharp(1996)}]{andersson_codon_1996}
Andersson, S.G. and Sharp, P.M. (1996).
\newblock \enquote{Codon usage and base composition in {Rickettsia}
  prowazekii.}
\newblock \emph{Journal of Molecular Evolution}, \textbf{42}(5): 525--536.

\bibitem[{Baldridge \emph{et~al.}(2014)Baldridge, Baldridge
  \emph{et~al.}}]{baldridge_proteomic_2014}
Baldridge, G.D., Baldridge, A.S., Witthuhn, B.A., Higgins, L., Markowski, T.W.,
  and Fallon, A.M. (2014).
\newblock \enquote{Proteomic profiling of a robust {Wolbachia} infection in an
  {Aedes} albopictus mosquito cell line.}
\newblock \emph{Molecular Microbiology}, \textbf{94}(3): 537--556.

\bibitem[{Beckmann and Fallon(2013)}]{beckmann_detection_2013}
Beckmann, J.F. and Fallon, A.M. (2013).
\newblock \enquote{Detection of the {Wolbachia} protein {WPIP}0282 in mosquito
  spermathecae: implications for cytoplasmic incompatibility.}
\newblock \emph{Insect Biochemistry and Molecular Biology}, \textbf{43}(9):
  867--878.

\bibitem[{Benjamini and Hochberg(1995)}]{benjamini_controlling_1995}
Benjamini, Y. and Hochberg, Y. (1995).
\newblock \enquote{Controlling the {False} {Discovery} {Rate}: {A} {Practical}
  and {Powerful} {Approach} to {Multiple} {Testing}.}
\newblock \emph{Journal of the Royal Statistical Society. Series B
  (Methodological)}, \textbf{57}(1): 289--300.

\bibitem[{Bennuru \emph{et~al.}(2011)Bennuru, Meng
  \emph{et~al.}}]{bennuru_stage-specific_2011}
Bennuru, S., Meng, Z., Ribeiro, J.M.C., Semnani, R.T., Ghedin, E., Chan, K.,
  Lucas, D.A., Veenstra, T.D., and Nutman, T.B. (2011).
\newblock \enquote{Stage-specific proteomic expression patterns of the human
  filarial parasite {Brugia} malayi and its endosymbiont {Wolbachia}.}
\newblock \emph{Proceedings of the National Academy of Sciences},
  \textbf{108}(23): 9649--9654.

\bibitem[{Bordenstein \emph{et~al.}(2006)Bordenstein, Marshall
  \emph{et~al.}}]{bordenstein_tripartite_2006}
Bordenstein, S.R., Marshall, M.L., Fry, A.J., Kim, U., and Wernegreen, J.J.
  (2006).
\newblock \enquote{The {Tripartite} {Associations} between {Bacteriophage},
  {Wolbachia}, and {Arthropods}.}
\newblock \emph{PLoS Pathog}, \textbf{2}(5): e43.

\bibitem[{Brizuela \emph{et~al.}(1994)Brizuela, Elfring
  \emph{et~al.}}]{brizuela_genetic_1994}
Brizuela, B.J., Elfring, L., Ballard, J., Tamkun, J.W., and Kennison, J.A.
  (1994).
\newblock \enquote{Genetic analysis of the brahma gene of {Drosophila}
  melanogaster and polytene chromosome subdivisions 72ab.}
\newblock \emph{Genetics}, \textbf{137}(3): 803--13.

\bibitem[{Brown \emph{et~al.}(2014)Brown, Boley
  \emph{et~al.}}]{brown_diversity_2014}
Brown, J.B., Boley, N., Eisman, R., May, G.E., Stoiber, M.H., Duff, M.O.,
  Booth, B.W., Wen, J., Park, S., Suzuki, A.M., Wan, K.H., Yu, C., Zhang, D.,
  Carlson, J.W., Cherbas, L., Eads, B.D., Miller, D., Mockaitis, K., Roberts,
  J., Davis, C.A., \emph{et~al.} (2014).
\newblock \enquote{Diversity and dynamics of the {Drosophila} transcriptome.}
\newblock \emph{Nature}, \textbf{512}(7515): 393--399.

\bibitem[{Caturegli \emph{et~al.}(2000)Caturegli, Asanovich
  \emph{et~al.}}]{caturegli_anka:_2000}
Caturegli, P., Asanovich, K.M., Walls, J.J., Bakken, J.S., Madigan, J.E.,
  Popov, V.L., and Dumler, J.S. (2000).
\newblock \enquote{{ankA}: an {Ehrlichia} phagocytophila group gene encoding a
  cytoplasmic protein antigen with ankyrin repeats.}
\newblock \emph{Infection and Immunity}, \textbf{68}(9): 5277--5283.

\bibitem[{Celniker \emph{et~al.}(2002)Celniker, Wheeler
  \emph{et~al.}}]{celniker_finishing_2002}
Celniker, S.E., Wheeler, D.A., Kronmiller, B., Carlson, J.W., Halpern, A.,
  Patel, S., Adams, M., Champe, M., Dugan, S.P., Frise, E., Hodgson, A.,
  George, R.A., Hoskins, R.A., Laverty, T., Muzny, D.M., Nelson, C.R., Pacleb,
  J.M., Park, S., Pfeiffer, B.D., Richards, S., \emph{et~al.} (2002).
\newblock \enquote{Finishing a whole-genome shotgun: release 3 of the
  {Drosophila} melanogaster euchromatic genome sequence.}
\newblock \emph{Genome Biol}, \textbf{3}(12): RESEARCH0079.

\bibitem[{Chan and Lowe(2009)}]{chan_gtrnadb:_2009}
Chan, P.P. and Lowe, T.M. (2009).
\newblock \enquote{{GtRNAdb}: a database of transfer {RNA} genes detected in
  genomic sequence.}
\newblock \emph{Nucleic Acids Research}, \textbf{37}(suppl 1): D93--D97.

\bibitem[{Christensen \emph{et~al.}(2001)Christensen, Mikkelsen
  \emph{et~al.}}]{christensen_rele_2001}
Christensen, S.K., Mikkelsen, M., Pedersen, K., and Gerdes, K. (2001).
\newblock \enquote{{RelE}, a global inhibitor of translation, is activated
  during nutritional stress.}
\newblock \emph{Proceedings of the National Academy of Sciences},
  \textbf{98}(25): 14328--14333.

\bibitem[{Chrostek \emph{et~al.}(2013)Chrostek, Marialva
  \emph{et~al.}}]{chrostek_wolbachia_2013}
Chrostek, E., Marialva, M.S.P., Esteves, S.S., Weinert, L.A., Martinez, J.,
  Jiggins, F.M., and Teixeira, L. (2013).
\newblock \enquote{Wolbachia {Variants} {Induce} {Differential} {Protection} to
  {Viruses} in {Drosophila} melanogaster: {A} {Phenotypic} and {Phylogenomic}
  {Analysis}.}
\newblock \emph{PLoS Genet}, \textbf{9}(12): e1003896.

\bibitem[{Chrostek and Teixeira(2015)}]{chrostek_mutualism_2015}
Chrostek, E. and Teixeira, L. (2015).
\newblock \enquote{Mutualism {Breakdown} by {Amplification} of {Wolbachia}
  {Genes}.}
\newblock \emph{PLoS Biol}, \textbf{13}(2): e1002065.

\bibitem[{Clark \emph{et~al.}(2005)Clark, Anderson
  \emph{et~al.}}]{clark_widespread_2005}
Clark, M.E., Anderson, C.L., Cande, J., and Karr, T.L. (2005).
\newblock \enquote{Widespread prevalence of wolbachia in laboratory stocks and
  the implications for {Drosophila} research.}
\newblock \emph{Genetics}, \textbf{170}(4): 1667--1675.

\bibitem[{Costanzo \emph{et~al.}(2008)Costanzo, Nicoloff
  \emph{et~al.}}]{costanzo_ppgpp_2008}
Costanzo, A., Nicoloff, H., Barchinger, S.E., Banta, A.B., Gourse, R.L., and
  Ades, S.E. (2008).
\newblock \enquote{{ppGpp} and {DksA} likely regulate the activity of the
  extracytoplasmic stress factor σ{E} in {Escherichia} coli by both direct and
  indirect mechanisms.}
\newblock \emph{Molecular Microbiology}, \textbf{67}(3): 619--632.

\bibitem[{Cunningham \emph{et~al.}(2015)Cunningham, Amode
  \emph{et~al.}}]{cunningham_ensembl_2015}
Cunningham, F., Amode, M.R., Barrell, D., Beal, K., Billis, K., Brent, S.,
  Carvalho-Silva, D., Clapham, P., Coates, G., Fitzgerald, S., Gil, L., Girón,
  C.G., Gordon, L., Hourlier, T., Hunt, S.E., Janacek, S.H., Johnson, N.,
  Juettemann, T., Kähäri, A.K., Keenan, S., \emph{et~al.} (2015).
\newblock \enquote{Ensembl 2015.}
\newblock \emph{Nucleic Acids Research}, \textbf{43}(Database issue):
  D662--669.

\bibitem[{Dale and Moran(2006)}]{dale_molecular_2006}
Dale, C. and Moran, N.A. (2006).
\newblock \enquote{Molecular {Interactions} between {Bacterial} {Symbionts} and
  {Their} {Hosts}.}
\newblock \emph{Cell}, \textbf{126}(3): 453--465.

\bibitem[{Dale \emph{et~al.}(2002)Dale, Plague \emph{et~al.}}]{dale_type_2002}
Dale, C., Plague, G.R., Wang, B., Ochman, H., and Moran, N.A. (2002).
\newblock \enquote{Type {III} secretion systems and the evolution of
  mutualistic endosymbiosis.}
\newblock \emph{Proceedings of the National Academy of Sciences of the United
  States of America}, \textbf{99}(19): 12397--12402.

\bibitem[{Darby \emph{et~al.}(2012)Darby, Armstrong
  \emph{et~al.}}]{darby_analysis_2012}
Darby, A.C., Armstrong, S.D., Bah, G.S., Kaur, G., Hughes, M.A., Kay, S.M.,
  Koldkjaer, P., Rainbow, L., Radford, A.D., Blaxter, M.L., Tanya, V.N., Trees,
  A.J., Cordaux, R., Wastling, J.M., and Makepeace, B.L. (2012).
\newblock \enquote{Analysis of gene expression from the {Wolbachia} genome of a
  filarial nematode supports both metabolic and defensive roles within the
  symbiosis.}
\newblock \emph{Genome Research}, \textbf{22}(12): 2467--2477.

\bibitem[{Darby \emph{et~al.}(2014)Darby, Christina~Gill
  \emph{et~al.}}]{darby_integrated_2014}
Darby, A.C., Christina~Gill, A., Armstrong, S.D., Hartley, C.S., Xia, D.,
  Wastling, J.M., and Makepeace, B.L. (2014).
\newblock \enquote{Integrated transcriptomic and proteomic analysis of the
  global response of {Wolbachia} to doxycycline-induced stress.}
\newblock \emph{The ISME Journal}, \textbf{8}(4): 925--937.

\bibitem[{de~Jonge \emph{et~al.}(2007)de~Jonge, Fehrmann
  \emph{et~al.}}]{de_jonge_evidence_2007}
de~Jonge, H.J.M., Fehrmann, R.S.N., de~Bont, E.S.J.M., Hofstra, R.M.W.,
  Gerbens, F., Kamps, W.A., de~Vries, E.G.E., van~der Zee, A.G.J., te~Meerman,
  G.J., and ter Elst, A. (2007).
\newblock \enquote{Evidence {Based} {Selection} of {Housekeeping} {Genes}.}
\newblock \emph{PLoS ONE}, \textbf{2}(9): e898.

\bibitem[{Dehal \emph{et~al.}(2010)Dehal, Joachimiak
  \emph{et~al.}}]{dehal_microbesonline:_2010}
Dehal, P.S., Joachimiak, M.P., Price, M.N., Bates, J.T., Baumohl, J.K.,
  Chivian, D., Friedland, G.D., Huang, K.H., Keller, K., Novichkov, P.S.,
  Dubchak, I.L., Alm, E.J., and Arkin, A.P. (2010).
\newblock \enquote{{MicrobesOnline}: an integrated portal for comparative and
  functional genomics.}
\newblock \emph{Nucleic Acids Research}, \textbf{38}(Database issue):
  D396--400.

\bibitem[{Duff \emph{et~al.}(2015)Duff, Olson
  \emph{et~al.}}]{duff_genome-wide_2015}
Duff, M.O., Olson, S., Wei, X., Garrett, S.C., Osman, A., Bolisetty, M.,
  Plocik, A., Celniker, S.E., and Graveley, B.R. (2015).
\newblock \enquote{Genome-wide identification of zero nucleotide recursive
  splicing in {Drosophila}.}
\newblock \emph{Nature}, \textbf{521}(7552): 376--379.

\bibitem[{Duron \emph{et~al.}(2007)Duron, Boureux
  \emph{et~al.}}]{duron_variability_2007}
Duron, O., Boureux, A., Echaubard, P., Berthomieu, A., Berticat, C., Fort, P.,
  and Weill, M. (2007).
\newblock \enquote{Variability and expression of ankyrin domain genes in
  {Wolbachia} variants infecting the mosquito {Culex} pipiens.}
\newblock \emph{Journal of Bacteriology}, \textbf{189}(12): 4442--4448.

\bibitem[{Fares \emph{et~al.}(2002)Fares, Barrio
  \emph{et~al.}}]{fares_evolution_2002}
Fares, M.A., Barrio, E., Sabater-Muñoz, B., and Moya, A. (2002).
\newblock \enquote{The evolution of the heat-shock protein {GroEL} from
  {Buchnera}, the primary endosymbiont of aphids, is governed by positive
  selection.}
\newblock \emph{Molecular Biology and Evolution}, \textbf{19}(7): 1162--1170.

\bibitem[{Finn \emph{et~al.}(2014)Finn, Bateman
  \emph{et~al.}}]{finn_pfam:_2014}
Finn, R.D., Bateman, A., Clements, J., Coggill, P., Eberhardt, R.Y., Eddy,
  S.R., Heger, A., Hetherington, K., Holm, L., Mistry, J., Sonnhammer, E.L.L.,
  Tate, J., and Punta, M. (2014).
\newblock \enquote{Pfam: the protein families database.}
\newblock \emph{Nucleic Acids Research}, \textbf{42}(Database issue):
  D222--230.

\bibitem[{Foster \emph{et~al.}(2005)Foster, Ganatra
  \emph{et~al.}}]{foster_wolbachia_2005}
Foster, J., Ganatra, M., Kamal, I., Ware, J., Makarova, K., Ivanova, N.,
  Bhattacharyya, A., Kapatral, V., Kumar, S., Posfai, J., Vincze, T., Ingram,
  J., Moran, L., Lapidus, A., Omelchenko, M., Kyrpides, N., Ghedin, E., Wang,
  S., Goltsman, E., Joukov, V., \emph{et~al.} (2005).
\newblock \enquote{The {Wolbachia} genome of {Brugia} malayi: endosymbiont
  evolution within a human pathogenic nematode.}
\newblock \emph{PLoS biology}, \textbf{3}(4): e121.

\bibitem[{Fujii \emph{et~al.}(2004)Fujii, Kubo
  \emph{et~al.}}]{fujii_isolation_2004}
Fujii, Y., Kubo, T., Ishikawa, H., and Sasaki, T. (2004).
\newblock \enquote{Isolation and characterization of the bacteriophage {WO}
  from {Wolbachia}, an arthropod endosymbiont.}
\newblock \emph{Biochemical and Biophysical Research Communications},
  \textbf{317}(4): 1183--1188.

\bibitem[{Gavotte \emph{et~al.}(2004)Gavotte, Vavre
  \emph{et~al.}}]{gavotte_diversity_2004}
Gavotte, L., Vavre, F., Henri, H., Ravallec, M., Stouthamer, R., and
  Boulétreau, M. (2004).
\newblock \enquote{Diversity, distribution and specificity of {WO} phage
  infection in {Wolbachia} of four insect species.}
\newblock \emph{Insect Molecular Biology}, \textbf{13}(2): 147--153.

\bibitem[{Gillespie \emph{et~al.}(2012)Gillespie, Joardar
  \emph{et~al.}}]{gillespie_rickettsia_2012}
Gillespie, J.J., Joardar, V., Williams, K.P., Driscoll, T., Hostetler, J.B.,
  Nordberg, E., Shukla, M., Walenz, B., Hill, C.A., Nene, V.M., Azad, A.F.,
  Sobral, B.W., and Caler, E. (2012).
\newblock \enquote{A {Rickettsia} {Genome} {Overrun} by {Mobile} {Genetic}
  {Elements} {Provides} {Insight} into the {Acquisition} of {Genes}
  {Characteristic} of an {Obligate} {Intracellular} {Lifestyle}.}
\newblock \emph{Journal of Bacteriology}, \textbf{194}(2): 376--394.

\bibitem[{Graveley \emph{et~al.}(2011)Graveley, Brooks
  \emph{et~al.}}]{graveley_developmental_2011}
Graveley, B.R., Brooks, A.N., Carlson, J.W., Duff, M.O., Landolin, J.M., Yang,
  L., Artieri, C.G., van Baren, M.J., Boley, N., Booth, B.W., Brown, J.B.,
  Cherbas, L., Davis, C.A., Dobin, A., Li, R., Lin, W., Malone, J.H.,
  Mattiuzzo, N.R., Miller, D., Sturgill, D., \emph{et~al.} (2011).
\newblock \enquote{The developmental transcriptome of {Drosophila}
  melanogaster.}
\newblock \emph{Nature}, \textbf{471}(7339): 473--479.

\bibitem[{Hansen and Degnan(2014)}]{hansen_widespread_2014}
Hansen, A.K. and Degnan, P.H. (2014).
\newblock \enquote{Widespread expression of conserved small {RNAs} in small
  symbiont genomes.}
\newblock \emph{The ISME Journal}, \textbf{8}(12): 2490--2502.

\bibitem[{Herbeck \emph{et~al.}(2003)Herbeck, Wall
  \emph{et~al.}}]{herbeck_gene_2003}
Herbeck, J.T., Wall, D.P., and Wernegreen, J.J. (2003).
\newblock \enquote{Gene expression level influences amino acid usage, but not
  codon usage, in the tsetse fly endosymbiont {Wigglesworthia}.}
\newblock \emph{Microbiology}, \textbf{149}(9): 2585--2596.

\bibitem[{Hoffmann \emph{et~al.}(1998)Hoffmann, Hercus
  \emph{et~al.}}]{hoffmann_population_1998}
Hoffmann, A.A., Hercus, M., and Dagher, H. (1998).
\newblock \enquote{Population dynamics of the {Wolbachia} infection causing
  cytoplasmic incompatibility in {Drosophila} melanogaster.}
\newblock \emph{Genetics}, \textbf{148}(1): 221--231.

\bibitem[{Hoffmann(1988)}]{hoffmann_partial_1988}
Hoffmann, A.A. (1988).
\newblock \enquote{Partial cytoplasmic incompatibility between two {Australian}
  populations of {Drosophila} melanogaster.}
\newblock \emph{Entomologia Experimentalis et Applicata}, \textbf{48}(1):
  61--67.

\bibitem[{Hoskins \emph{et~al.}(2015)Hoskins, Carlson
  \emph{et~al.}}]{hoskins_release_2015}
Hoskins, R.A., Carlson, J.W., Wan, K.H., Park, S., Mendez, I., Galle, S.E.,
  Booth, B.W., Pfeiffer, B.D., George, R.A., Svirskas, R., Krzywinski, M.,
  Schein, J., Accardo, M.C., Damia, E., Messina, G., Méndez-Lago, M.,
  de~Pablos, B., Demakova, O.V., Andreyeva, E.N., Boldyreva, L.V.,
  \emph{et~al.} (2015).
\newblock \enquote{The {Release} 6 reference sequence of the {Drosophila}
  melanogaster genome.}
\newblock \emph{Genome Research}, \textbf{25}(3): 445--458.

\bibitem[{Huang \emph{et~al.}(2014)Huang, Massouras
  \emph{et~al.}}]{huang_natural_2014}
Huang, W., Massouras, A., Inoue, Y., Peiffer, J., Ràmia, M., Tarone, A.M.,
  Turlapati, L., Zichner, T., Zhu, D., Lyman, R.F., Magwire, M.M., Blankenburg,
  K., Carbone, M.A., Chang, K., Ellis, L.L., Fernandez, S., Han, Y., Highnam,
  G., Hjelmen, C.E., Jack, J.R., \emph{et~al.} (2014).
\newblock \enquote{Natural variation in genome architecture among 205
  {Drosophila} melanogaster {Genetic} {Reference} {Panel} lines.}
\newblock \emph{Genome Research}, \textbf{24}(7): 1193--1208.

\bibitem[{Ishmael \emph{et~al.}(2009)Ishmael, Hotopp
  \emph{et~al.}}]{ishmael_extensive_2009}
Ishmael, N., Hotopp, J.C.D., Ioannidis, P., Biber, S., Sakamoto, J., Siozios,
  S., Nene, V., Werren, J., Bourtzis, K., Bordenstein, S.R., and Tettelin, H.
  (2009).
\newblock \enquote{Extensive genomic diversity of closely related {Wolbachia}
  strains.}
\newblock \emph{Microbiology}, \textbf{155}(Pt 7): 2211--2222.

\bibitem[{Iturbe-Ormaetxe \emph{et~al.}(2005)Iturbe-Ormaetxe, Burke
  \emph{et~al.}}]{iturbe-ormaetxe_distribution_2005}
Iturbe-Ormaetxe, I., Burke, G.R., Riegler, M., and O'Neill, S.L. (2005).
\newblock \enquote{Distribution, expression, and motif variability of ankyrin
  domain genes in {Wolbachia} pipientis.}
\newblock \emph{Journal of Bacteriology}, \textbf{187}(15): 5136--5145.

\bibitem[{Kent and Bordenstein(2010)}]{kent_phage_2010}
Kent, B.N. and Bordenstein, S.R. (2010).
\newblock \enquote{Phage {WO} of {Wolbachia}: lambda of the endosymbiont
  world.}
\newblock \emph{Trends in Microbiology}, \textbf{18}(4): 173--181.

\bibitem[{Kent \emph{et~al.}(2011)Kent, Funkhouser
  \emph{et~al.}}]{kent_evolutionary_2011}
Kent, B.N., Funkhouser, L.J., Setia, S., and Bordenstein, S.R. (2011).
\newblock \enquote{Evolutionary {Genomics} of a {Temperate} {Bacteriophage} in
  an {Obligate} {Intracellular} {Bacteria} ({Wolbachia}).}
\newblock \emph{PLoS ONE}, \textbf{6}(9): e24984.

\bibitem[{Kersey \emph{et~al.}(2014)Kersey, Allen
  \emph{et~al.}}]{kersey_ensembl_2014}
Kersey, P.J., Allen, J.E., Christensen, M., Davis, P., Falin, L.J.,
  Grabmueller, C., Hughes, D.S.T., Humphrey, J., Kerhornou, A., Khobova, J.,
  Langridge, N., McDowall, M.D., Maheswari, U., Maslen, G., Nuhn, M., Ong,
  C.K., Paulini, M., Pedro, H., Toneva, I., Tuli, M.A., \emph{et~al.} (2014).
\newblock \enquote{Ensembl {Genomes} 2013: scaling up access to genome-wide
  data.}
\newblock \emph{Nucleic Acids Research}, \textbf{42}(Database issue):
  D546--552.

\bibitem[{Klasson \emph{et~al.}(2008)Klasson, Walker
  \emph{et~al.}}]{klasson_genome_2008}
Klasson, L., Walker, T., Sebaihia, M., Sanders, M.J., Quail, M.A., Lord, A.,
  Sanders, S., Earl, J., O'Neill, S.L., Thomson, N., Sinkins, S.P., and
  Parkhill, J. (2008).
\newblock \enquote{Genome evolution of {Wolbachia} strain {wPip} from the
  {Culex} pipiens group.}
\newblock \emph{Molecular Biology and Evolution}, \textbf{25}(9): 1877--87.

\bibitem[{Klasson \emph{et~al.}(2009)Klasson, Westberg
  \emph{et~al.}}]{klasson_mosaic_2009}
Klasson, L., Westberg, J., Sapountzis, P., Näslund, K., Lutnaes, Y., Darby,
  A.C., Veneti, Z., Chen, L., Braig, H.R., Garrett, R., Bourtzis, K., and
  Andersson, S.G. (2009).
\newblock \enquote{The mosaic genome structure of the {Wolbachia} {wRi} strain
  infecting {Drosophila} simulans.}
\newblock \emph{Proceedings of the National Academy of Sciences of the United
  States of America}, \textbf{106}(14): 5725--5730.

\bibitem[{Li and Dewey(2011)}]{li_rsem:_2011}
Li, B. and Dewey, C.N. (2011).
\newblock \enquote{{RSEM}: accurate transcript quantification from {RNA}-{Seq}
  data with or without a reference genome.}
\newblock \emph{BMC Bioinformatics}, \textbf{12}(1): 323.

\bibitem[{Li(2013)}]{li_aligning_2013}
Li, H. (2013).
\newblock \enquote{Aligning sequence reads, clone sequences and assembly
  contigs with {BWA}-{MEM}.}
\newblock \emph{arXiv}, page 1303.3997.

\bibitem[{Li \emph{et~al.}(2009)Li, Handsaker \emph{et~al.}}]{li_sequence_2009}
Li, H., Handsaker, B., Wysoker, A., Fennell, T., Ruan, J., Homer, N., Marth,
  G., Abecasis, G., and Durbin, R. (2009).
\newblock \enquote{The {Sequence} {Alignment}/{Map} format and {SAMtools}.}
\newblock \emph{Bioinformatics}, \textbf{25}(16): 2078--2079.

\bibitem[{Lin \emph{et~al.}(2007)Lin, den Dulk-Ras
  \emph{et~al.}}]{lin_anaplasma_2007}
Lin, M., den Dulk-Ras, A., Hooykaas, P.J.J., and Rikihisa, Y. (2007).
\newblock \enquote{Anaplasma phagocytophilum {AnkA} secreted by type {IV}
  secretion system is tyrosine phosphorylated by {Abl}-1 to facilitate
  infection.}
\newblock \emph{Cellular Microbiology}, \textbf{9}(11): 2644--2657.

\bibitem[{Luck \emph{et~al.}(2014)Luck, Evans
  \emph{et~al.}}]{luck_concurrent_2014}
Luck, A.N., Evans, C.C., Riggs, M.D., Foster, J.M., Moorhead, A.R., Slatko,
  B.E., and Michalski, M.L. (2014).
\newblock \enquote{Concurrent transcriptional profiling of {Dirofilaria}
  immitis and its {Wolbachia} endosymbiont throughout the nematode life cycle
  reveals coordinated gene expression.}
\newblock \emph{BMC Genomics}, \textbf{15}(1): 1041.

\bibitem[{Mao \emph{et~al.}(2014)Mao, Ma \emph{et~al.}}]{mao_door_2014}
Mao, X., Ma, Q., Zhou, C., Chen, X., Zhang, H., Yang, J., Mao, F., Lai, W., and
  Xu, Y. (2014).
\newblock \enquote{{DOOR} 2.0: presenting operons and their functions through
  dynamic and integrated views.}
\newblock \emph{Nucleic Acids Research}, \textbf{42}(Database issue):
  D654--659.

\bibitem[{Masui \emph{et~al.}(2001)Masui, Kuroiwa
  \emph{et~al.}}]{masui_bacteriophage_2001}
Masui, S., Kuroiwa, H., Sasaki, T., Inui, M., Kuroiwa, T., and Ishikawa, H.
  (2001).
\newblock \enquote{Bacteriophage {WO} and virus-like particles in {Wolbachia},
  an endosymbiont of arthropods.}
\newblock \emph{Biochem Biophys Res Commun}, \textbf{283}(5): 1099--104.

\bibitem[{Mayoral \emph{et~al.}(2014)Mayoral, Hussain
  \emph{et~al.}}]{mayoral_wolbachia_2014}
Mayoral, J.G., Hussain, M., Joubert, D.A., Iturbe-Ormaetxe, I., O’Neill,
  S.L., and Asgari, S. (2014).
\newblock \enquote{Wolbachia small noncoding {RNAs} and their role in
  cross-kingdom communications.}
\newblock \emph{Proceedings of the National Academy of Sciences},
  \textbf{111}(52): 18721--18726.

\bibitem[{McCarthy \emph{et~al.}(2012)McCarthy, Chen
  \emph{et~al.}}]{mccarthy_differential_2012}
McCarthy, D.J., Chen, Y., and Smyth, G.K. (2012).
\newblock \enquote{Differential expression analysis of multifactor {RNA}-{Seq}
  experiments with respect to biological variation.}
\newblock \emph{Nucleic Acids Research}, \textbf{40}(10): 4288--4297.

\bibitem[{Metcalf \emph{et~al.}(2014)Metcalf, Jo
  \emph{et~al.}}]{metcalf_recent_2014}
Metcalf, J.A., Jo, M., Bordenstein, S.R., Jaenike, J., and Bordenstein, S.R.
  (2014).
\newblock \enquote{Recent genome reduction of {Wolbachia} in {Drosophila}
  recens targets phage {WO} and narrows candidates for reproductive
  parasitism.}
\newblock \emph{PeerJ}, \textbf{2}: e529.

\bibitem[{Min and Benzer(1997)}]{min_wolbachia_1997}
Min, K.T. and Benzer, S. (1997).
\newblock \enquote{Wolbachia, normally a symbiont of {Drosophila}, can be
  virulent, causing degeneration and early death.}
\newblock \emph{Proceedings of the National Academy of Sciences of the United
  States of America}, \textbf{94}(20): 10792--10796.

\bibitem[{Moran(1996)}]{moran_accelerated_1996}
Moran, N.A. (1996).
\newblock \enquote{Accelerated evolution and {Muller}'s rachet in endosymbiotic
  bacteria.}
\newblock \emph{Proceedings of the National Academy of Sciences of the United
  States of America}, \textbf{93}(7): 2873--2878.

\bibitem[{Moran and Degnan(2006)}]{moran_functional_2006}
Moran, N.A. and Degnan, P.H. (2006).
\newblock \enquote{Functional genomics of {Buchnera} and the ecology of aphid
  hosts.}
\newblock \emph{Molecular Ecology}, \textbf{15}(5): 1251--1261.

\bibitem[{Moran \emph{et~al.}(2005)Moran, Dunbar
  \emph{et~al.}}]{moran_regulation_2005}
Moran, N.A., Dunbar, H.E., and Wilcox, J.L. (2005).
\newblock \enquote{Regulation of transcription in a reduced bacterial genome:
  nutrient-provisioning genes of the obligate symbiont {Buchnera} aphidicola.}
\newblock \emph{Journal of Bacteriology}, \textbf{187}(12): 4229--4237.

\bibitem[{Moran \emph{et~al.}(2008)Moran, McCutcheon
  \emph{et~al.}}]{moran_genomics_2008}
Moran, N.A., McCutcheon, J.P., and Nakabachi, A. (2008).
\newblock \enquote{Genomics and evolution of heritable bacterial symbionts.}
\newblock \emph{Annual review of genetics}, \textbf{42}(1): 165--190.

\bibitem[{O’Brien \emph{et~al.}(2015)O’Brien, Lindsay
  \emph{et~al.}}]{obrien_legionella_2015}
O’Brien, K.M., Lindsay, E.L., and Starai, V.J. (2015).
\newblock \enquote{The {Legionella} pneumophila {Effector} {Protein}, {LegC}7,
  {Alters} {Yeast} {Endosomal} {Trafficking}.}
\newblock \emph{PLoS ONE}, \textbf{10}(2).

\bibitem[{Palacios and Wernegreen(2002)}]{palacios_strong_2002}
Palacios, C. and Wernegreen, J.J. (2002).
\newblock \enquote{A {Strong} {Effect} of {AT} {Mutational} {Bias} on {Amino}
  {Acid} {Usage} in {Buchnera} is {Mitigated} at {High}-{Expression} {Genes}.}
\newblock \emph{Molecular Biology and Evolution}, \textbf{19}(9): 1575--1584.

\bibitem[{Pan \emph{et~al.}(2008)Pan, Lührmann
  \emph{et~al.}}]{pan_ankyrin_2008}
Pan, X., Lührmann, A., Satoh, A., Laskowski-Arce, M.A., and Roy, C.R. (2008).
\newblock \enquote{Ankyrin {Repeat} {Proteins} {Comprise} a {Diverse} {Family}
  of {Bacterial} {Type} {IV} {Effectors}.}
\newblock \emph{Science}, \textbf{320}(5883): 1651--1654.

\bibitem[{Papafotiou \emph{et~al.}(2011)Papafotiou, Oehler
  \emph{et~al.}}]{papafotiou_regulation_2011}
Papafotiou, G., Oehler, S., Savakis, C., and Bourtzis, K. (2011).
\newblock \enquote{Regulation of {Wolbachia} ankyrin domain encoding genes in
  {Drosophila} gonads.}
\newblock \emph{Research in Microbiology}, \textbf{162}(8): 764--772.

\bibitem[{Paul \emph{et~al.}(2005)Paul, Berkmen \emph{et~al.}}]{paul_dksa_2005}
Paul, B.J., Berkmen, M.B., and Gourse, R.L. (2005).
\newblock \enquote{{DksA} potentiates direct activation of amino acid promoters
  by {ppGpp}.}
\newblock \emph{Proceedings of the National Academy of Sciences},
  \textbf{102}(22): 7823--7828.

\bibitem[{Pedersen \emph{et~al.}(2003)Pedersen, Zavialov
  \emph{et~al.}}]{pedersen_bacterial_2003}
Pedersen, K., Zavialov, A.V., Pavlov, M.Y., Elf, J., Gerdes, K., and Ehrenberg,
  M. (2003).
\newblock \enquote{The bacterial toxin {RelE} displays codon-specific cleavage
  of {mRNAs} in the ribosomal {A} site.}
\newblock \emph{Cell}, \textbf{112}(1): 131--140.

\bibitem[{Perkins \emph{et~al.}(2009)Perkins, Kingsley
  \emph{et~al.}}]{perkins_strand-specific_2009}
Perkins, T.T., Kingsley, R.A., Fookes, M.C., Gardner, P.P., James, K.D., Yu,
  L., Assefa, S.A., He, M., Croucher, N.J., Pickard, D.J., Maskell, D.J.,
  Parkhill, J., Choudhary, J., Thomson, N.R., and Dougan, G. (2009).
\newblock \enquote{A {Strand}-{Specific} {RNA}–{Seq} {Analysis} of the
  {Transcriptome} of the {Typhoid} {Bacillus} {Salmonella} {Typhi}.}
\newblock \emph{PLoS Genet}, \textbf{5}(7): e1000569.

\bibitem[{Pertea \emph{et~al.}(2009)Pertea, Ayanbule
  \emph{et~al.}}]{pertea_operondb:_2009}
Pertea, M., Ayanbule, K., Smedinghoff, M., and Salzberg, S.L. (2009).
\newblock \enquote{{OperonDB}: a comprehensive database of predicted operons in
  microbial genomes.}
\newblock \emph{Nucleic Acids Research}, \textbf{37}(suppl 1): D479--D482.

\bibitem[{Pfaffl(2001)}]{pfaffl_new_2001}
Pfaffl, M.W. (2001).
\newblock \enquote{A new mathematical model for relative quantification in
  real-time {RT}–{PCR}.}
\newblock \emph{Nucleic Acids Research}, \textbf{29}(9): e45--e45.

\bibitem[{Pinto \emph{et~al.}(2013)Pinto, Stainton
  \emph{et~al.}}]{pinto_transcriptional_2013}
Pinto, S.B., Stainton, K., Harris, S., Kambris, Z., Sutton, E.R., Bonsall,
  M.B., Parkhill, J., and Sinkins, S.P. (2013).
\newblock \enquote{Transcriptional {Regulation} of {Culex} pipiens {Mosquitoes}
  by {Wolbachia} {Influences} {Cytoplasmic} {Incompatibility}.}
\newblock \emph{PLoS Pathog}, \textbf{9}(10): e1003647.

\bibitem[{Poinsot \emph{et~al.}(2003)Poinsot, Charlat
  \emph{et~al.}}]{poinsot_mechanism_2003}
Poinsot, D., Charlat, S., and Merçot, H. (2003).
\newblock \enquote{On the mechanism of {Wolbachia}-induced cytoplasmic
  incompatibility: confronting the models with the facts.}
\newblock \emph{BioEssays}, \textbf{25}(3): 259--265.

\bibitem[{Quinlan and Hall(2010)}]{quinlan_bedtools:_2010}
Quinlan, A.R. and Hall, I.M. (2010).
\newblock \enquote{{BEDTools}: a flexible suite of utilities for comparing
  genomic features.}
\newblock \emph{Bioinformatics}, \textbf{26}(6): 841--842.

\bibitem[{Rao \emph{et~al.}(2012)Rao, Huang \emph{et~al.}}]{rao_effects_2012}
Rao, R.U., Huang, Y., Abubucker, S., Heinz, M., Crosby, S.D., Mitreva, M., and
  Weil, G.J. (2012).
\newblock \enquote{Effects of doxycycline on gene expression in {Wolbachia} and
  {Brugia} malayi adult female worms in vivo.}
\newblock \emph{Journal of Biomedical Science}, \textbf{19}: 21.

\bibitem[{Reymond \emph{et~al.}(2006)Reymond, Calevro
  \emph{et~al.}}]{reymond_different_2006}
Reymond, N., Calevro, F., Viñuelas, J., Morin, N., Rahbé, Y., Febvay, G.,
  Laugier, C., Douglas, A., Fayard, J.M., and Charles, H. (2006).
\newblock \enquote{Different levels of transcriptional regulation due to
  trophic constraints in the reduced genome of {Buchnera} aphidicola {APS}.}
\newblock \emph{Applied and Environmental Microbiology}, \textbf{72}(12):
  7760--7766.

\bibitem[{Reynolds and Hoffmann(2002)}]{reynolds_male_2002}
Reynolds, K.T. and Hoffmann, A.A. (2002).
\newblock \enquote{Male age, host effects and the weak expression or
  non-expression of cytoplasmic incompatibility in {Drosophila} strains
  infected by maternally transmitted {Wolbachia}.}
\newblock \emph{Genetical Research}, \textbf{80}(2): 79--87.

\bibitem[{Reynolds \emph{et~al.}(2003)Reynolds, Thomson
  \emph{et~al.}}]{reynolds_effects_2003}
Reynolds, K.T., Thomson, L.J., and Hoffmann, A.A. (2003).
\newblock \enquote{The effects of host age, host nuclear background and
  temperature on phenotypic effects of the virulent {Wolbachia} strain popcorn
  in {Drosophila} melanogaster.}
\newblock \emph{Genetics}, \textbf{164}(3): 1027--1034.

\bibitem[{Richardson \emph{et~al.}(2012)Richardson, Weinert
  \emph{et~al.}}]{richardson_population_2012}
Richardson, M.F., Weinert, L.A., Welch, J.J., Linheiro, R.S., Magwire, M.M.,
  Jiggins, F.M., and Bergman, C.M. (2012).
\newblock \enquote{Population {Genomics} of the {Wolbachia} {Endosymbiont} in
  {Drosophila} melanogaster.}
\newblock \emph{PLoS Genet}, \textbf{8}(12): e1003129.

\bibitem[{Rispe \emph{et~al.}(2004)Rispe, Delmotte
  \emph{et~al.}}]{rispe_mutational_2004}
Rispe, C., Delmotte, F., Ham, R.C.H.J.v., and Moya, A. (2004).
\newblock \enquote{Mutational and {Selective} {Pressures} on {Codon} and
  {Amino} {Acid} {Usage} in {Buchnera}, {Endosymbiotic} {Bacteria} of
  {Aphids}.}
\newblock \emph{Genome Research}, \textbf{14}(1): 44--53.

\bibitem[{Robinson \emph{et~al.}(2010)Robinson, McCarthy
  \emph{et~al.}}]{robinson_edger:_2010}
Robinson, M.D., McCarthy, D.J., and Smyth, G.K. (2010).
\newblock \enquote{{edgeR}: a {Bioconductor} package for differential
  expression analysis of digital gene expression data.}
\newblock \emph{Bioinformatics}, \textbf{26}(1): 139--140.

\bibitem[{Robinson and Oshlack(2010)}]{robinson_scaling_2010}
Robinson, M.D. and Oshlack, A. (2010).
\newblock \enquote{A scaling normalization method for differential expression
  analysis of {RNA}-seq data.}
\newblock \emph{Genome Biology}, \textbf{11}(3): R25.

\bibitem[{Robinson and Smyth(2007)}]{robinson_moderated_2007}
Robinson, M.D. and Smyth, G.K. (2007).
\newblock \enquote{Moderated statistical tests for assessing differences in tag
  abundance.}
\newblock \emph{Bioinformatics (Oxford, England)}, \textbf{23}(21): 2881--2887.

\bibitem[{Robinson and Smyth(2008)}]{robinson_small-sample_2008}
Robinson, M.D. and Smyth, G.K. (2008).
\newblock \enquote{Small-sample estimation of negative binomial dispersion,
  with applications to {SAGE} data.}
\newblock \emph{Biostatistics (Oxford, England)}, \textbf{9}(2): 321--332.

\bibitem[{Salzberg \emph{et~al.}(2005)Salzberg, Hotopp
  \emph{et~al.}}]{salzberg_serendipitous_2005}
Salzberg, S.L., Hotopp, J.C.D., Delcher, A.L., Pop, M., Smith, D.R., Eisen,
  M.B., and Nelson, W.C. (2005).
\newblock \enquote{Serendipitous discovery of {Wolbachia} genomes in multiple
  {Drosophila} species.}
\newblock \emph{Genome Biol}, \textbf{6}(3): R23.

\bibitem[{Sanogo and Dobson(2006)}]{sanogo_wo_2006}
Sanogo, Y.O. and Dobson, S.L. (2006).
\newblock \enquote{{WO} bacteriophage transcription in {Wolbachia}-infected
  {Culex} pipiens.}
\newblock \emph{Insect Biochemistry and Molecular Biology}, \textbf{36}(1):
  80--85.

\bibitem[{Shao \emph{et~al.}(2011)Shao, Harrison
  \emph{et~al.}}]{shao_tadb:_2011}
Shao, Y., Harrison, E.M., Bi, D., Tai, C., He, X., Ou, H.Y., Rajakumar, K., and
  Deng, Z. (2011).
\newblock \enquote{{TADB}: a web-based resource for {Type} 2 toxin–antitoxin
  loci in bacteria and archaea.}
\newblock \emph{Nucleic Acids Research}, \textbf{39}(suppl 1): D606--D611.

\bibitem[{Si \emph{et~al.}(2014)Si, Liu \emph{et~al.}}]{si_model-based_2014}
Si, Y., Liu, P., Li, P., and Brutnell, T.P. (2014).
\newblock \enquote{Model-based clustering for {RNA}-seq data.}
\newblock \emph{Bioinformatics}, \textbf{30}(2): 197--205.

\bibitem[{Sinkins \emph{et~al.}(2005)Sinkins, Walker
  \emph{et~al.}}]{sinkins_wolbachia_2005}
Sinkins, S.P., Walker, T., Lynd, A.R., Steven, A.R., Makepeace, B.L., Godfray,
  H.C.J., and Parkhill, J. (2005).
\newblock \enquote{Wolbachia variability and host effects on crossing type in
  {Culex} mosquitoes.}
\newblock \emph{Nature}, \textbf{436}(7048): 257--260.

\bibitem[{Siozios \emph{et~al.}(2013)Siozios, Ioannidis
  \emph{et~al.}}]{siozios_diversity_2013}
Siozios, S., Ioannidis, P., Klasson, L., Andersson, S.G.E., Braig, H.R., and
  Bourtzis, K. (2013).
\newblock \enquote{The {Diversity} and {Evolution} of {Wolbachia} {Ankyrin}
  {Repeat} {Domain} {Genes}.}
\newblock \emph{PLoS ONE}, \textbf{8}(2): e55390.

\bibitem[{Sisko \emph{et~al.}(2006)Sisko, Spaeth
  \emph{et~al.}}]{sisko_multifunctional_2006}
Sisko, J.L., Spaeth, K., Kumar, Y., and Valdivia, R.H. (2006).
\newblock \enquote{Multifunctional analysis of {Chlamydia}-specific genes in a
  yeast expression system.}
\newblock \emph{Molecular Microbiology}, \textbf{60}(1): 51--66.

\bibitem[{Slatko \emph{et~al.}(2014)Slatko, Luck
  \emph{et~al.}}]{slatko_wolbachia_2014}
Slatko, B.E., Luck, A.N., Dobson, S.L., and Foster, J.M. (2014).
\newblock \enquote{Wolbachia endosymbionts and human disease control.}
\newblock \emph{Molecular and Biochemical Parasitology}, \textbf{195}(2):
  88--95.

\bibitem[{Stamatakis(2014)}]{stamatakis_raxml_2014}
Stamatakis, A. (2014).
\newblock \enquote{{RAxML} version 8: a tool for phylogenetic analysis and
  post-analysis of large phylogenies.}
\newblock \emph{Bioinformatics}, \textbf{30}(9): 1312--1313.

\bibitem[{Taboada \emph{et~al.}(2012)Taboada, Ciria
  \emph{et~al.}}]{taboada_proopdb:_2012}
Taboada, B., Ciria, R., Martinez-Guerrero, C.E., and Merino, E. (2012).
\newblock \enquote{{ProOpDB}: {Prokaryotic} {Operon} {DataBase}.}
\newblock \emph{Nucleic Acids Research}, \textbf{40}(D1): D627--D631.

\bibitem[{Team(2012)}]{r_development_core_team_r:_2012}
Team, R.D.C. (2012).
\newblock \emph{R: {A} {Language} and {Environment} for {Statistical}
  {Computing}}.
\newblock Vienna, Austria.
\newblock ISBN 3-900051-07-0.

\bibitem[{Van~Melderen and Saavedra
  De~Bast(2009)}]{van_melderen_bacterial_2009}
Van~Melderen, L. and Saavedra De~Bast, M. (2009).
\newblock \enquote{Bacterial {Toxin}–{Antitoxin} {Systems}: {More} {Than}
  {Selfish} {Entities}?}
\newblock \emph{PLoS Genet}, \textbf{5}(3): e1000437.

\bibitem[{Wang \emph{et~al.}(2014)Wang, Niu \emph{et~al.}}]{wang_large_2014}
Wang, G.H., Niu, L.M., Ma, G.C., Xiao, J.H., and Huang, D.W. (2014).
\newblock \enquote{Large proportion of genes in one cryptic {WO} prophage
  genome are actively and sex-specifically transcribed in a fig wasp species.}
\newblock \emph{BMC Genomics}, \textbf{15}(1): 893.

\bibitem[{Werren \emph{et~al.}(2008)Werren, Baldo
  \emph{et~al.}}]{werren_wolbachia:_2008}
Werren, J.H., Baldo, L., and Clark, M.E. (2008).
\newblock \enquote{Wolbachia: master manipulators of invertebrate biology.}
\newblock \emph{Nat Rev Microbiol}, \textbf{6}(10): 741--51.

\bibitem[{Wilcox \emph{et~al.}(2003)Wilcox, Dunbar
  \emph{et~al.}}]{wilcox_consequences_2003}
Wilcox, J.L., Dunbar, H.E., Wolfinger, R.D., and Moran, N.A. (2003).
\newblock \enquote{Consequences of reductive evolution for gene expression in
  an obligate endosymbiont.}
\newblock \emph{Molecular Microbiology}, \textbf{48}(6): 1491--1500.

\bibitem[{Woolfit \emph{et~al.}(2015)Woolfit, Algama
  \emph{et~al.}}]{woolfit_discovery_2015}
Woolfit, M., Algama, M., Keith, J.M., McGraw, E.A., and Popovici, J. (2015).
\newblock \enquote{Discovery of {Putative} {Small} {Non}-{Coding} {RNAs} from
  the {Obligate} {Intracellular} {Bacterium} {Wolbachia} pipientis.}
\newblock \emph{PLoS ONE}, \textbf{10}(3): e0118595.

\bibitem[{Woolfit \emph{et~al.}(2013)Woolfit, Iturbe-Ormaetxe
  \emph{et~al.}}]{woolfit_genomic_2013}
Woolfit, M., Iturbe-Ormaetxe, I., Brownlie, J.C., Walker, T., Riegler, M.,
  Seleznev, A., Popovici, J., Rances, E., Wee, B.A., Pavlides, J., Sullivan,
  M.J., Beatson, S.A., Lane, A., Sidhu, M., McMeniman, C.J., McGraw, E.A., and
  O'Neill, S.L. (2013).
\newblock \enquote{Genomic {Evolution} of the {Pathogenic} {Wolbachia}
  {Strain}, {wMelPop}.}
\newblock \emph{Genome Biology and Evolution}, \textbf{5}(11): 2189--2204.

\bibitem[{Wu \emph{et~al.}(2004)Wu, Sun \emph{et~al.}}]{wu_phylogenomics_2004}
Wu, M., Sun, L.V., Vamathevan, J., Riegler, M., Deboy, R., Brownlie, J.C.,
  McGraw, E.A., Martin, W., Esser, C., Ahmadinejad, N., Wiegand, C., Madupu,
  R., Beanan, M.J., Brinkac, L.M., Daugherty, S.C., Durkin, A.S., Kolonay,
  J.F., Nelson, W.C., Mohamoud, Y., Lee, P., \emph{et~al.} (2004).
\newblock \enquote{Phylogenomics of the {Reproductive} {Parasite} {Wolbachia}
  pipientis {wMel}: {A} {Streamlined} {Genome} {Overrun} by {Mobile} {Genetic}
  {Elements}.}
\newblock \emph{PLoS Biol}, \textbf{2}(3): e69.

\bibitem[{Yamada \emph{et~al.}(2007)Yamada, Floate
  \emph{et~al.}}]{yamada_male_2007}
Yamada, R., Floate, K.D., Riegler, M., and O'Neill, S.L. (2007).
\newblock \enquote{Male development time influences the strength of
  {Wolbachia}-induced cytoplasmic incompatibility expression in {Drosophila}
  melanogaster.}
\newblock \emph{Genetics}, \textbf{177}(2): 801--808.

\bibitem[{Yamaguchi \emph{et~al.}(2011)Yamaguchi, Park
  \emph{et~al.}}]{yamaguchi_toxin-antitoxin_2011}
Yamaguchi, Y., Park, J.H., and Inouye, M. (2011).
\newblock \enquote{Toxin-{Antitoxin} {Systems} in {Bacteria} and {Archaea}.}
\newblock \emph{Annual Review of Genetics}, \textbf{45}(1): 61--79.

\bibitem[{Zug and Hammerstein(2015)}]{zug_bad_2015}
Zug, R. and Hammerstein, P. (2015).
\newblock \enquote{Bad guys turned nice? {A} critical assessment of {Wolbachia}
  mutualisms in arthropod hosts.}
\newblock \emph{Biological Reviews}, \textbf{90}(1): 89--111.

\end{thebibliography}
\clearpage

\section*{Figures}

\clearpage

\section*{Supplemental Figures}
\label{sec:Supplemental Figures}
\setcounter{figure}{0}
\renewcommand{\figurename}{Supplemental Figure}

\begin{figure}[h!] \centering
\includegraphics[width=0.6\textwidth]{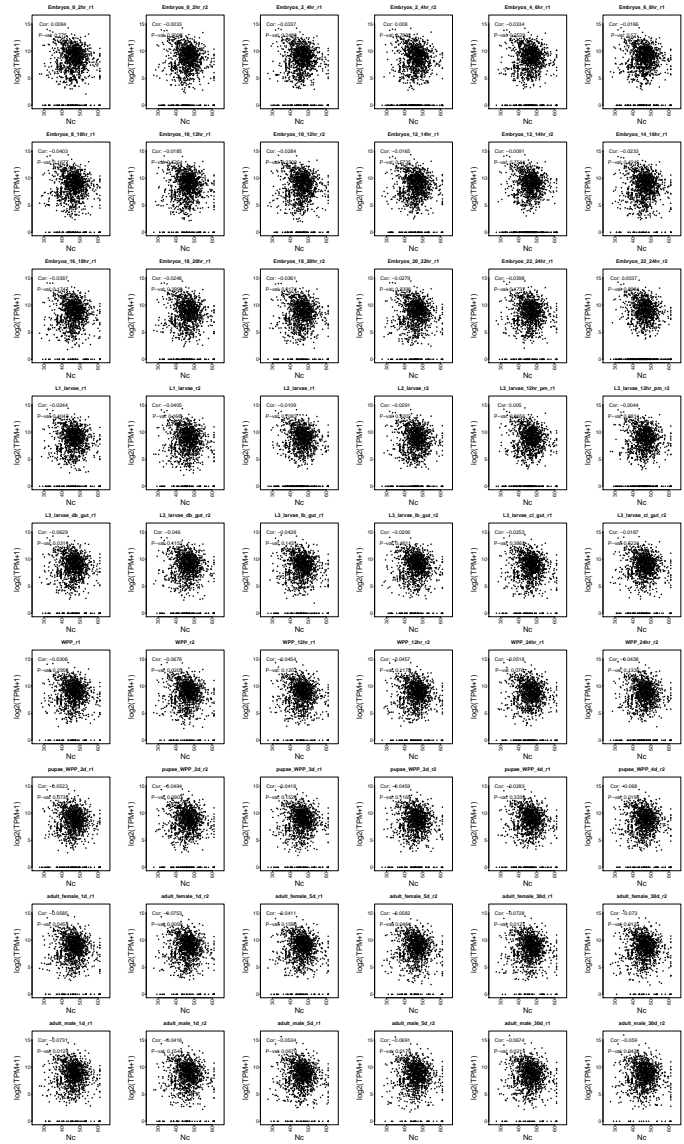}
\caption{{\bf Correlation between codon usage bias and expression level of \textit{Wolbachia} genes across different stages of the \textit{D. melanogaster} life cycle.} Pairwise scatterplots between  \textit{Wolbachia} gene expression levels (in units of TPM) and codon usage bias (measured as effective number of codons, $N_{c}$), with associated Pearson correlation coefficients. Each panel represents correlation of codon usage with a different sample in the modENCODE total RNA life cycle time course. A pseudocount of one was added to each gene's TPM before transforming to log2 scale.}
\label{fig:codon_bias} 
\end{figure}

\begin{figure}[ht] \centering
\includegraphics[width=0.85\textwidth]{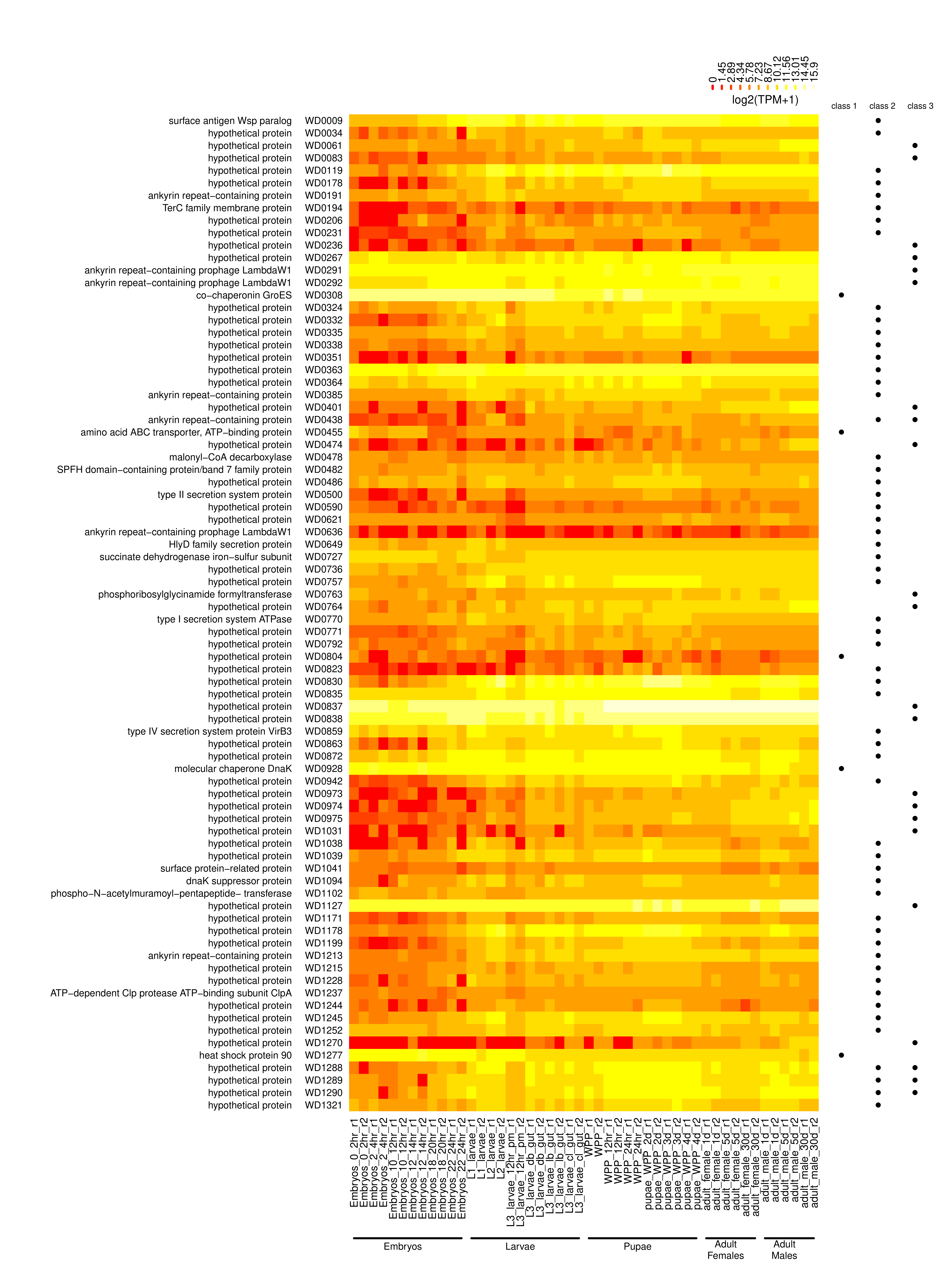}
\caption{{\bf Non-row-normalized expression levels for \textit{Wolbachia} genes that differ in expression across the \textit{D. melanogaster} life cycle.} Expression levels are visualized as a heatmap where each row represents a gene (ordered top-to-bottom by its position in the genome) and each cell represents expression in units of TPM. A pseudocount of one was added to each gene's TPM before transforming to log2 scale. Values with higher levels of expression are represented by yellow, and values with lower levels of expression are represented by red.  Gene names and identifiers are shown on the left. Membership in different dynamically-expressed gene classes is shown by dots on the right. Class 1 includes genes that show down-regulation after embryogenesis. Class 2 includes genes that show up-regulation after embryogenesis, with peaks of expression in larval and pupal stages. Class 3 includes genes that show up-regulation after embryogenesis, with peaks of expression in adult. Classification of gene sets is not mutually exclusive. Stages that lack biological replicates in the modENCODE total RNA-seq time course were not used in this analysis and are not shown here.}
\label{fig:heatmap_iso1DE_nonorm} 
\end{figure}

\begin{figure}[ht] \centering
\includegraphics[width=0.70\textwidth]{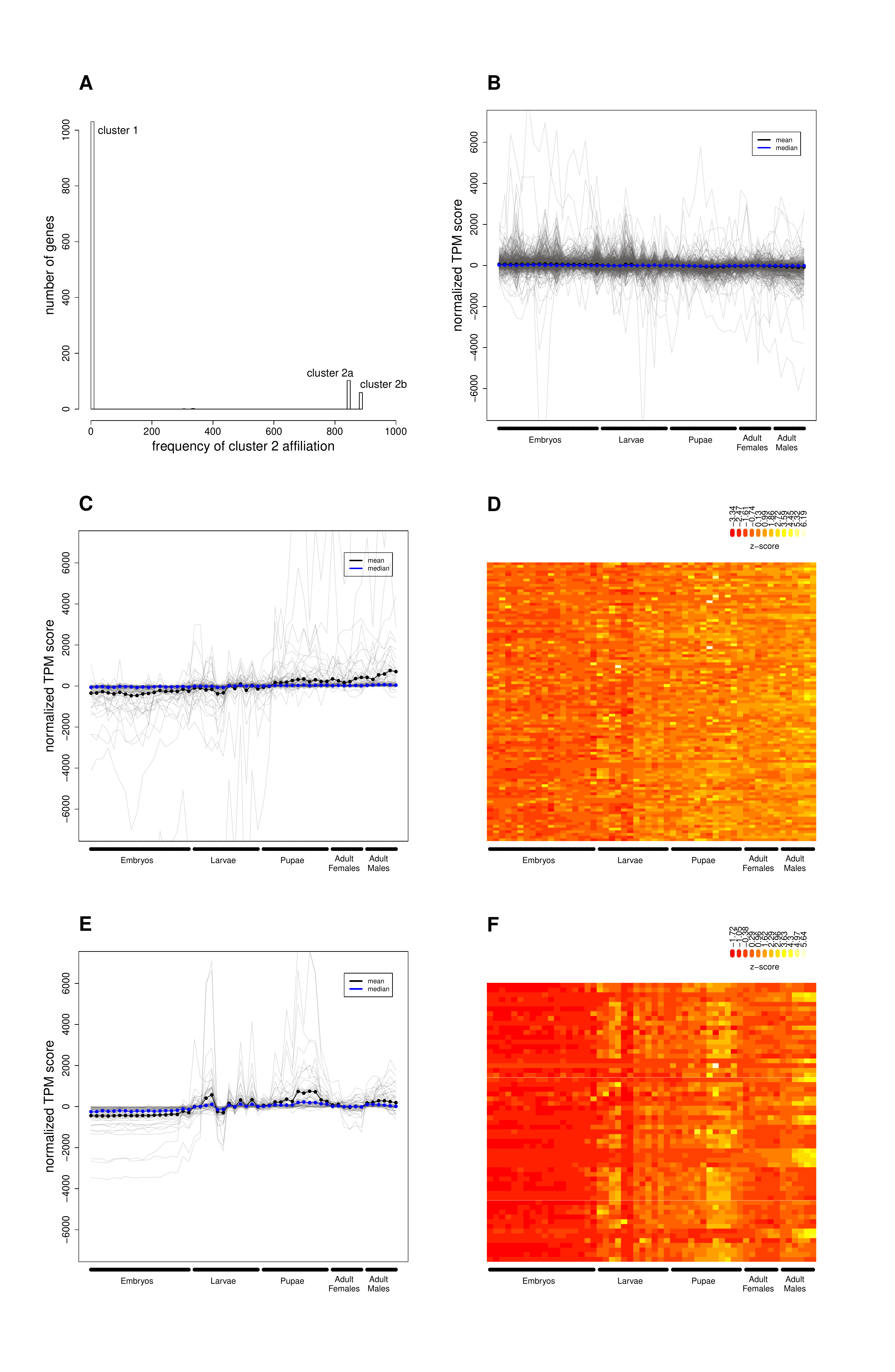} 
\caption{{\bf Clustering analysis of \textit{Wolbachia} gene expression in the modENCODE life cycle time course.} A. Histogram showing number of independent clustering runs (out of 1000) a gene was affiliated with the variable gene cluster 2. Three distinct peaks were observed, one for stably-expressed genes (cluster 1, n=\nbCone), and two peaks for variable genes that we denote cluster 2a (n=\nbCtwoa) and cluster 2b (n=\nbCtwob). B. Normalized expression profiles for genes in Cluster 1. C. Normalized expression profiles for genes in Cluster 2a. D. Heat map of row-normalized expression levels for genes in Cluster 2a. E. Normalized expression profiles for genes in Cluster 2b. F. Heat map of row-normalized expression levels for genes in Cluster 2b.   For panels B, C and E, TPMs for each gene at each stage are normalized by subtracting the mean TPM for that gene across different life cycle stages (shown in grey). The mean and median of normalized expression levels for all genes at each stage are show for each cluster in black and blue, respectively. For panels D and F, row-normalized expression levels are visualized as a heatmap where each row represents a gene (ordered top-to-bottom by its position in the genome), and each cell represents the relative expression level for a particular sample in terms of Z-scores (observed TPM minus row mean TPM, divided by the standard deviation of TPMs for that row). Values higher than row means are represented by yellow, and values lower than row means are represented by red.  The heatmap color scale differs in panels D and F.}
\label{fig:clustering} 
\end{figure}

\begin{figure}[ht] \centering
\includegraphics[width=1.0\textwidth]{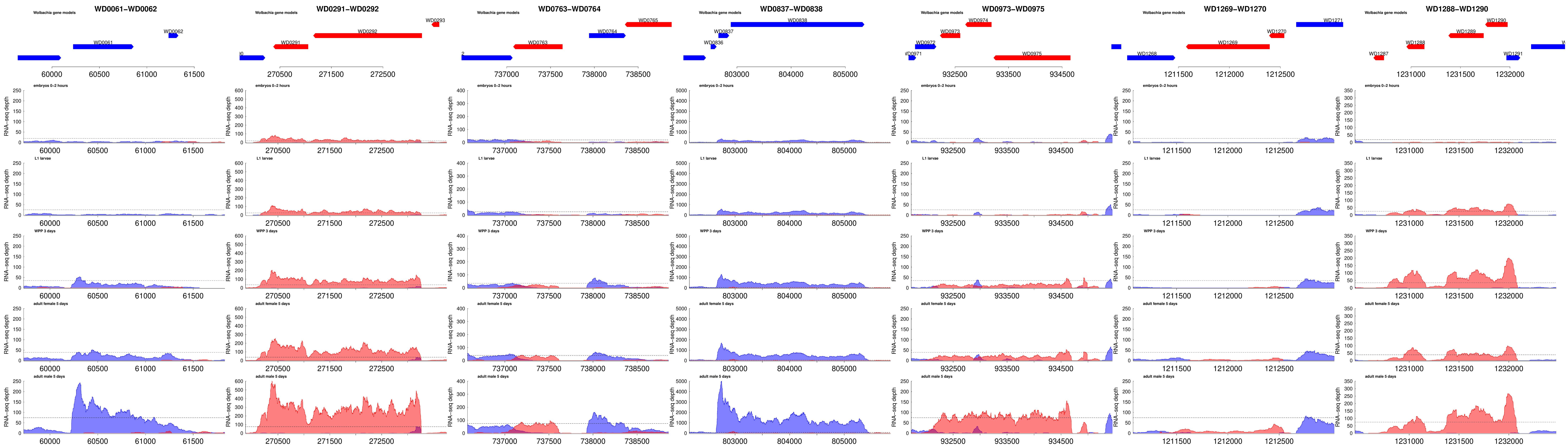}
\caption{{\bf \textit{Wolbachia} genes with sex-biased expression are often found in operons.} Wiggle plots of \textit{Wolbachia} expression levels for seven clusters of \textit{Wolbachia} genes with sex-biased expression. Gene models and RNA-seq coverage for each stage are shown for the forward and reverse strands in blue and red, respectively. Gene models and RNA-seq coverage for each stage are shown for the forward and reverse strands in blue and red, respectively. RNA-seq plots are shown on the same absolute y-axis scale. To provide an internal normalization factor for comparison across samples, mean coverage of the stably-expressed Wsp/WD1063 gene (not shown in this interval) divided by twenty is depicted by the dashed line in each panel.  Six out of seven clusters (WD0061--WD0062, WD0291--WD0292, WD0837--WD0838, WD0973--WD0975, WD1269--WD1270 and WD1288--WD1290) were confirmed as operons based on contiguous mapping of RNA-seq reads. The third cluster depicted contains two divergently transcribed genes (WD0763--WD0764) that are not co-transcribed as an operon.}
\label{fig:operons} 
\end{figure}

\begin{figure}[ht] \centering
\includegraphics[width=0.15\textwidth]{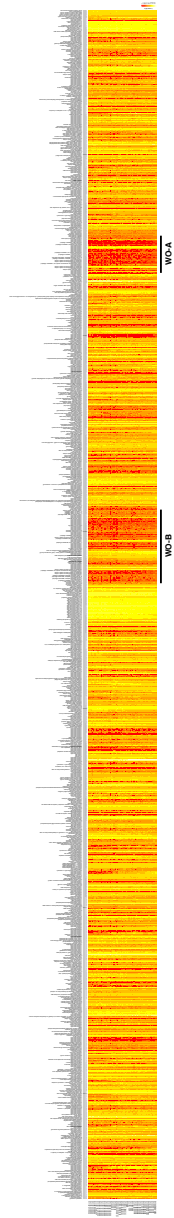}
\caption{{\bf Expression profiles of all \textit{Wolbachia} genes across the \textit{D. melanogaster} life cycle.}  The two major prophage regions -- WO-A and WO-B -- in the \textit{w}Mel genome are indicated. Expression levels are visualized as a heatmap where each row represents a gene (ordered top-to-bottom by its position in the genome) and each cell represents expression in units of TPM. A pseudocount of one was added to each gene's TPM before transforming to log2 scale. Values with higher levels of expression are represented by yellow, and values with lower levels of expression are represented by red.  Gene names and identifiers are shown on the left. All stages including those that lack biological replicates in the modENCODE time course are shown here.}
\label{fig:genome_heatmap} 
\end{figure}

\begin{figure}[ht] \centering
\includegraphics[width=1.0\textwidth]{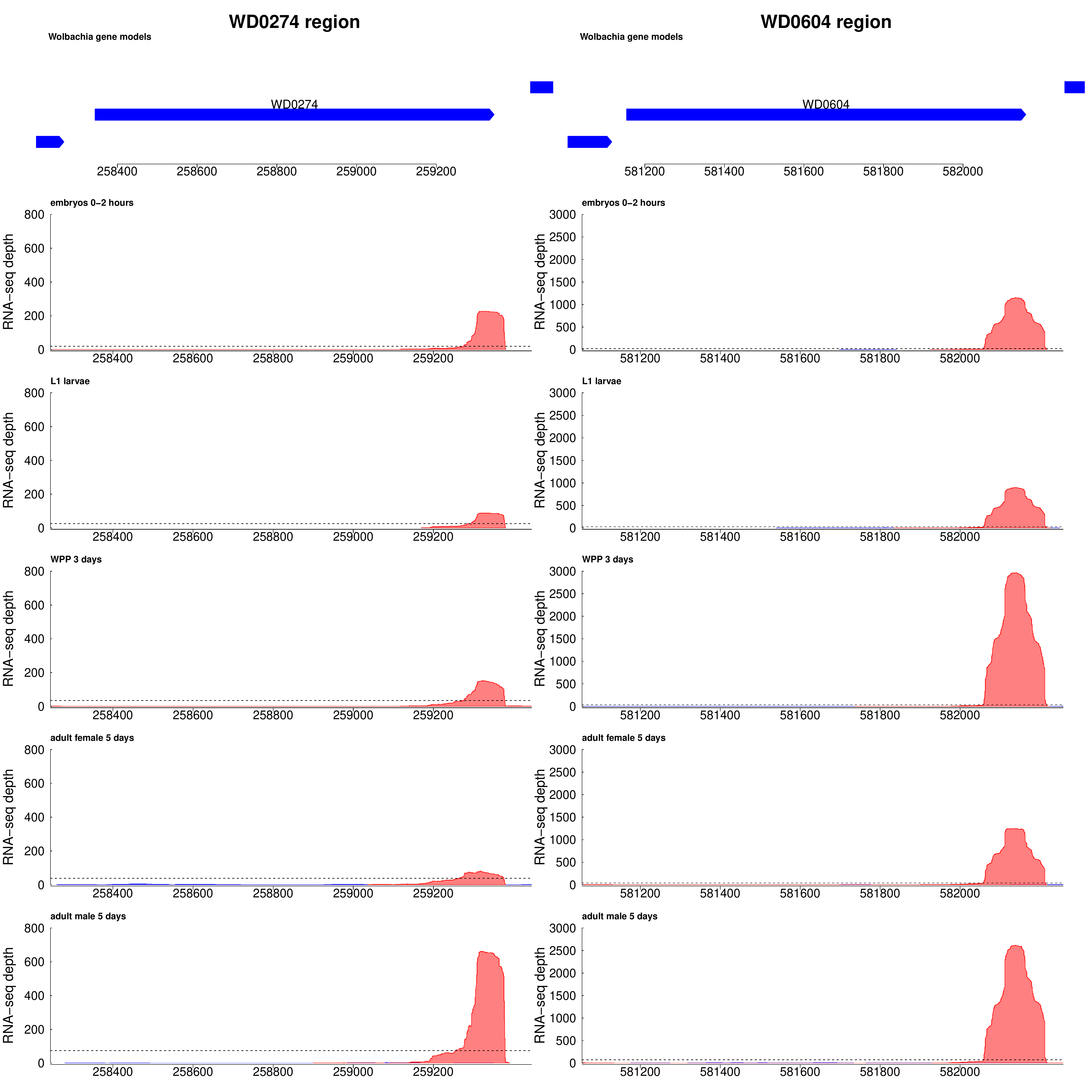}
\caption{{\bf Putative anti-sense noncoding RNAs in the \textit{Wolbachia} WO-A and WO-B regions.} Wiggle plots of \textit{Wolbachia} expression levels for two highly expressed putative anti-sense noncoding RNA genes that overlap the 3' ends of the major phage capsid genes of both WO-A (WD0274) and WO-B (WD0604). Gene models and RNA-seq coverage for each stage are shown for the forward and reverse strands in blue and red, respectively. RNA-seq plots are shown on the same absolute y-axis scale. To provide provide an internal normalization factor for comparison across samples, mean coverage of the stably-expressed Wsp/WD1063 gene (not shown in this interval) divided by twenty is depicted by the dashed line in each panel.  These transcribed regions are the most highly expressed sequences in both of the WO-A and WO-B regions, and are found in conserved locations of paralogs with divergent sequences.}
\label{fig:ncRNAs} 
\end{figure}

\clearpage

\section*{Supplemental Files}
\label{sec:Supplemental Files}
\begin{enumerate}

\item \label{itm:supplemental_file_1.tsv} A .tsv file with SRA IDs, summary of mapped reads for \textit{D. melanogaster} and \textit{Wolbachia}, number of expressed genes (defined as genes with non-zero TPM or genes with $\geq 2$ mapped reads per gene), mean TPM for the sample (same for all samples, inverse of gene number time one million), standard deviation of TPM for the sample.

\item \label{itm:supplemental_file_2.tsv} A .tsv file with gene IDs, coordinates, number of reads (from read 2 of paired end data) mapping to the sense strand in each sample (\_r1 = replicate 1, \_r2 = replicate 2), estimated TPM for each sample, number of runs found in cluster 2, cluster assignment, adjusted p-value in life-cycle GLM, log2 fold-change in life-cycle GLM vs embryo 0-2 hours, maximum fold change between any two stages in life-cycle GLM, adjusted p-values and log2 fold change for pairwise exact tests between male and female samples at 1, 5 and 30 days, gene name, annotated gene product, effective number of codons, GC content, number of homologs in \textit{w}Mel, wRi, \textit{w}Pip-Pel, and \textit{w}Bm genomes.

\item \label{itm:supplemental_file_3.tsv} A .tsv file with results of GLM for RT-qPCR analysis in ISO1 and w1118. P-values reported are not adjusted for multiple testing, and thus $\alpha$-levels for significance were set at 0.001.

\item \label{itm:supplemental_file_4.tsv} A .tsv file with PCR primers for RT-qPCR experiments.

\end{enumerate}
\end{document}